\documentclass[10pt,journal,compsoc,twoside]{IEEEtran}

\usepackage[nocompress]{cite}
\usepackage{xcolor}
\usepackage{amsmath}
\usepackage{amssymb}
\usepackage{amsfonts}
\usepackage{graphicx}
\usepackage{textcomp, dirtytalk}
\usepackage[section]{algorithm}
\usepackage{algcompatible}
\usepackage[nomessages]{fp}
\usepackage[hidelinks]{hyperref}
\usepackage{ragged2e}

\DeclareMathOperator{\NOT}{NOT}
\DeclareMathOperator{\OR}{OR}
\DeclareMathOperator{\AND}{AND}
\DeclareMathOperator{\XOR}{XOR}
\DeclareMathOperator{\XNOR}{XNOR}
\DeclareMathOperator{\FA}{FA}
\DeclareMathOperator{\Maj}{Maj}
\DeclareMathOperator{\mux}{mux}
\DeclareMathOperator{\Id}{Id}

\usepackage{soul}
\newcommand{\revision}[1]{#1}

\begin{document}

\title{AritPIM: High-Throughput In-Memory Arithmetic}

\author{Orian~Leitersdorf,~\IEEEmembership{Student Member,~IEEE,}
        Dean~Leitersdorf,
        Jonathan~Gal,
        Mor~Dahan,
        Ronny~Ronen,~\IEEEmembership{Fellow,~IEEE,}
        and~Shahar~Kvatinsky,~\IEEEmembership{Senior Member,~IEEE}%
\IEEEcompsocitemizethanks{\IEEEcompsocthanksitem O. Leitersdorf, D. Leitersdorf, J. Gal, M. Dahan, R. Ronen, and S. Kvatinsky are with the Technion -- Israel Institute of Technology, Haifa, Israel. 
E-mail: orianl@campus.technion.ac.il; leitersdorf@cs.technion.ac.il;
jonathan.gal@campus.technion.ac.il;
mor.dahan@campus.technion.ac.il;
ronny.ronen@technion.ac.il; 
shahar@ee.technion.ac.il.}%
}

\markboth{IEEE Transactions on Emerging Topics in Computing}{Leitersdorf \MakeLowercase{\textit{et al.}}: High-Throughput In-Memory Arithmetic}

\IEEEtitleabstractindextext{%

\begin{abstract}
\justifying
Digital processing-in-memory (PIM) architectures are rapidly emerging to overcome the memory-wall bottleneck by integrating logic within memory elements. Such architectures provide vast computational power within the memory itself in the form of parallel bitwise logic operations. We develop novel \emph{algorithmic} techniques for PIM that, combined with new perspectives on computer arithmetic, extend this bitwise parallelism to the four fundamental arithmetic operations (addition, subtraction, multiplication, and division), for both fixed-point and floating-point numbers, and using both bit-serial and bit-parallel approaches. We propose a state-of-the-art suite of arithmetic algorithms, demonstrating the first algorithm in the literature of digital PIM for a majority of cases -- including cases previously considered impossible for digital PIM, such as floating-point addition. Through a case study on memristive PIM, we compare the proposed algorithms to an NVIDIA RTX 3070 GPU and demonstrate significant throughput and energy improvements.
\end{abstract}

\begin{IEEEkeywords}
Digital processing-in-memory (PIM), parallel computation, arithmetic, fixed-point arithmetic, floating-point arithmetic.
\end{IEEEkeywords}}

\maketitle

\IEEEpubid{\begin{minipage}{\textwidth}\ \\[12pt] \centering \copyright 2023 IEEE. Personal use of this material is permitted.  Permission from IEEE must be obtained for all other uses, in any current or future media, including reprinting/republishing this material for advertising or promotional purposes, creating new collective works, for resale or redistribution to servers or lists, or reuse of any copyrighted component of this work in other works.
\end{minipage}} 

\IEEEdisplaynontitleabstractindextext

\IEEEraisesectionheading{\section{Introduction}\label{sec:introduction}}

\IEEEpubidadjcol

\IEEEPARstart{E}{merging} processing-in-memory (PIM) systems attempt to overcome the memory-wall bottleneck by rethinking one of the core principles of computing systems: the separation of storage and logic units. This separation has been followed since the introduction of the von Neumann architecture in the 1940s, when computing systems were primarily utilized for \emph{serial} program execution. Yet, the recent emergence of data-intensive applications requires \emph{parallel high-throughput} execution, causing the separation to become a massive bottleneck known as the \emph{memory wall}~\cite{Horowitz2014}. Therefore, PIM integrates logic within the memory itself to bypass the bandwidth-limited memory interface and enable massive in-memory computational parallelism~\cite{StatefulLogicReview}. 

\begin{figure}[!t]
\centering 
\includegraphics[width=\linewidth, trim={0cm, 0.2cm, 0cm, 0cm}]{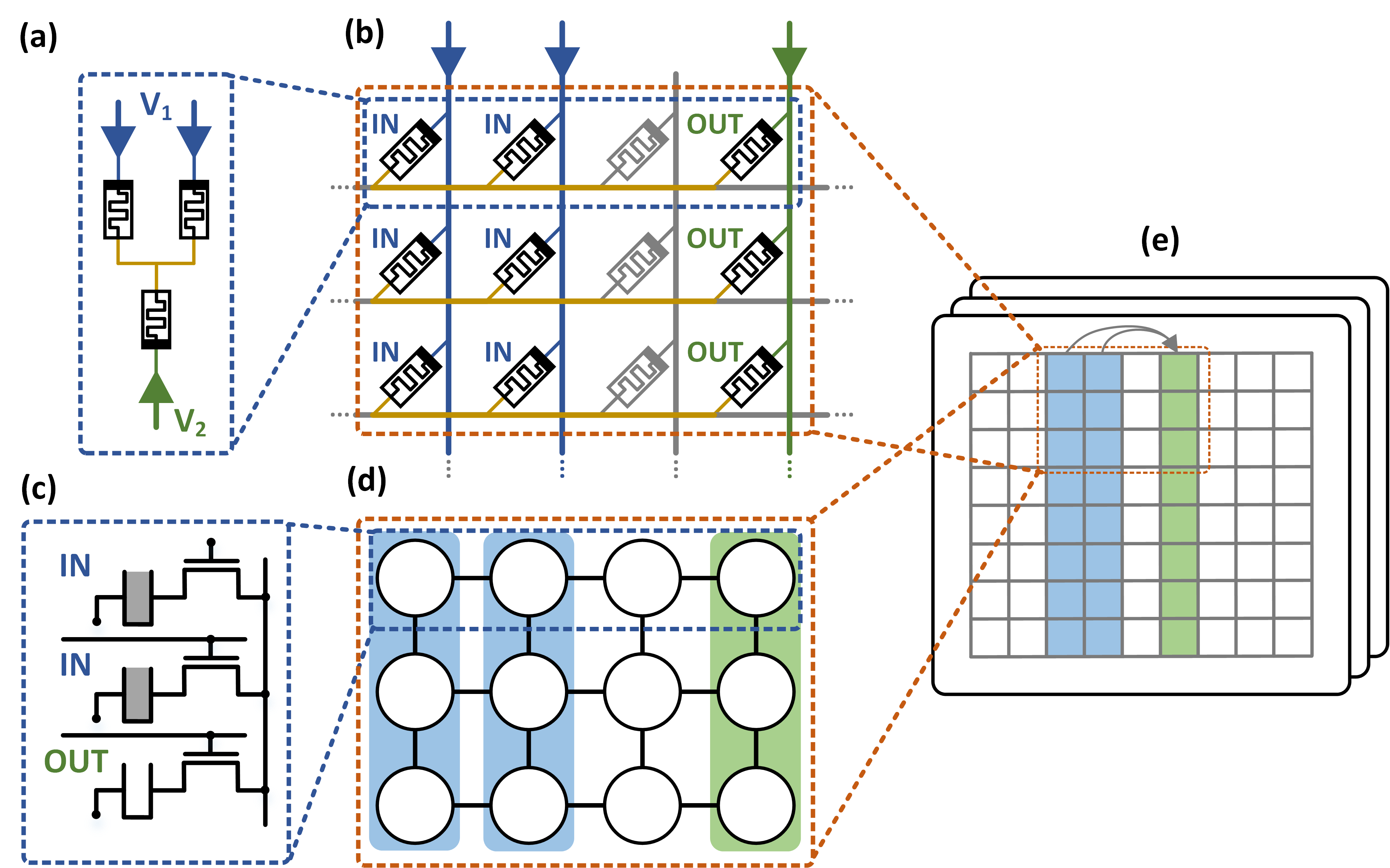}
\caption{Examples of PIM technologies for (a, b) memristive~\cite{Nishil, FELIX, CRAM, 8697377} and (c, d) DRAM~\cite{Ambit} memories. These follow (e) an abstract model of arbitrary bitwise column operations in $O(1)$ latency.}
\label{fig:PIM} 
\vspace{-10pt}
\end{figure}

PIM architectures supplement the traditional read/write memory interface with logic~\cite{StatefulLogicReview}. This enables the CPU to request that the memory perform \emph{vectored} logic on data stored within the memory without transferring the data through the interface, thereby significantly reducing the load on CPU-memory communication. Early proposals for PIM~\cite{IntelligentRAM} involved integrating logic circuits \emph{near} the memory (e.g., in the same chip), yet this still requires a fundamental need for data-transfer between an area dedicated for computation and an area dedicated for storage~\cite{StatefulLogicReview}. Conversely, recent proposals~\cite{Nishil, RACER, SIMDRAM, ComputeDRAM, potter2012associative, GSI, ComputeSRAM} perform \emph{digital} logic using the same physical devices that store binary information. By performing the logic exactly where the information is stored, data-transfer is effectively dwarfed~\cite{StatefulLogicReview}. These include works that exploit content-addressable-memories (CAMs)~\cite{GSI, potter2012associative, ComputeSRAM} to selectively apply write operations according to data stored in the memory (serving as inputs), and works that design logic gates from the underlying circuits connecting the memory devices~\cite{Nishil, RACER, SIMDRAM, ComputeDRAM}. Several proposals for PIM architectures, such as memristive (Figure~\ref{fig:PIM}(a,b))~\cite{Nishil, RACER, FELIX, IMPLY, CRAM, 8697377} and DRAM-based (Figure~\ref{fig:PIM}(c,d))~\cite{Ambit, ComputeDRAM, SIMDRAM}, have essentially converged to a single abstract computational model that presents highly unique algorithmic capabilities (Figure~\ref{fig:PIM}(e)). First, consider the memory as a collection of $m$ binary matrices, called \emph{arrays}, each of dimension $r \times c$. A \emph{bitwise} operation can be performed on columns of an array, and in parallel across all arrays, in a single cycle ($O(1)$ latency). For example, the bit-wise NOR~\cite{Nishil} of any two columns can be computed and stored in a third column, all in a single cycle. This is possible as the logic is performed in a \emph{distributed} fashion amongst the physical elements within arrays (with shared instructions), so there is no centralized computing unit that may cause a bottleneck. This attains massive parallelism for \emph{bitwise} operations that bypass the memory interface.

\begin{figure*}[!t]
\centering 
\includegraphics[width=\linewidth, trim={0cm, 0.3cm, 0cm, 0cm}]{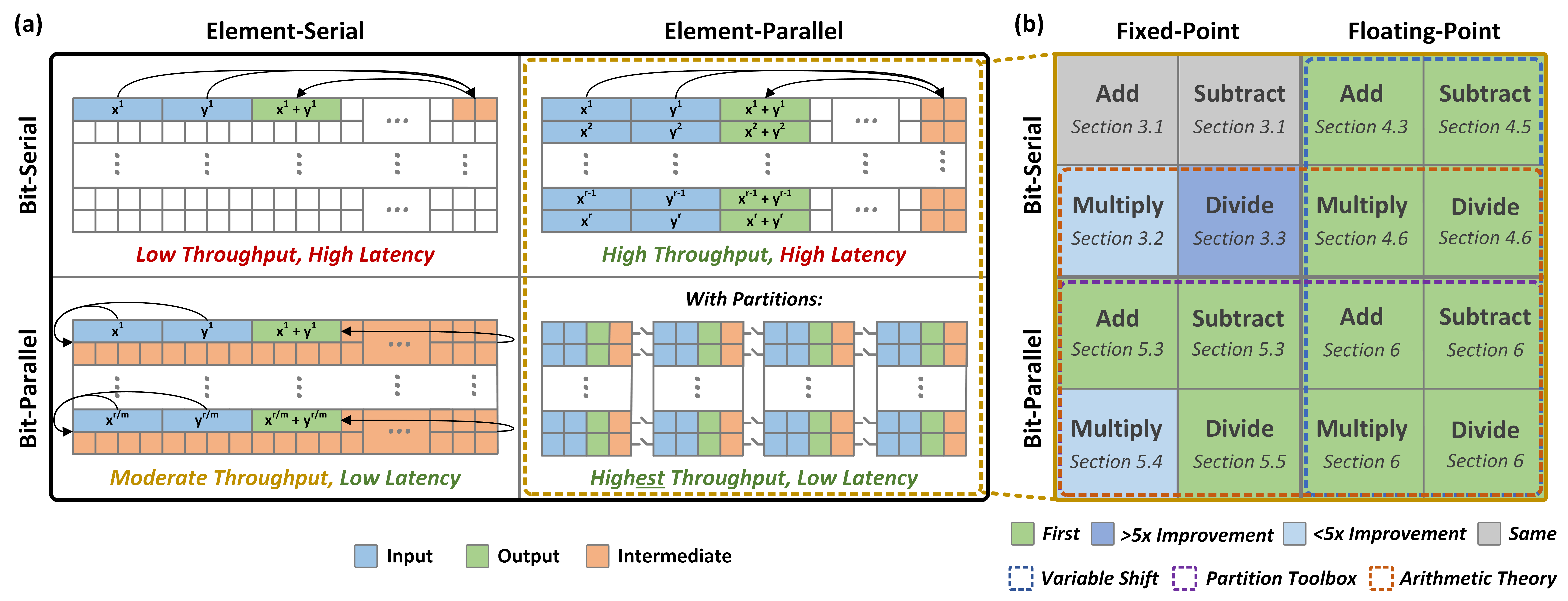}
\caption{(a) The various approaches to in-memory arithmetic developed in recent years. (b) The impact of this paper seen through the algorithms for the foundational arithmetic functions on both fixed-point and floating-point numbers, and with both bit-serial element-parallel and bit-parallel element-parallel approaches. The dashed rectangles highlight the three novel methods developed, and the algorithms which they effect.}
\label{fig:Approaches} 
\vspace{-10pt}
\end{figure*}

Expanding the massive bitwise parallelism to large-scale applications requires a strong foundation for fundamental arithmetic operations (addition, subtraction, multiplication, and division), for both fixed-point and floating-point numbers. While theoretically a functionally-complete set of logic gates can perform any function, the data layout plays a crucial role in the efficient utilization of PIM. Figure~\ref{fig:Approaches}(a) presents an overview of the approaches developed in recent years; we focus without loss of generality on a single array (as computation is in parallel across all arrays regardless). We describe the approaches through $N$-bit integer vector addition, where $x^i$ refers to the $i^{th}$ element in the vector:
\begin{enumerate}
    \item \emph{Bit-Serial Element-Serial:} Two inputs, $x^1$ and $y^1$, are stored within a single row of an array, and basic logic gates (e.g., NOR) \emph{serially} construct an $N$-bit adder within that row (utilizing intermediate cells for temporary results). This has \emph{low throughput} as only one addition is performed per array and \emph{high latency} as the gates run serially; thus, this approach is \emph{typically not used}. Note that column parallelism (constant-time column operations) is not utilized. 
    \item \emph{Bit-Parallel Element-Serial~\cite{UltraEfficient, WallaceTree, SIMPLE, CRAM, 8416761}:} \revision{$m$ rows are utilized} as intermediate space to perform multiple parallel gates for the same $N$-bit adder by utilizing column parallelism, \revision{thereby enabling $r/m$ adders per array}. This provides \emph{low latency} when the function possesses parallelism among the gates (typically best applicable to multiplication~\cite{WallaceTree}), yet \revision{possesses} \emph{\revision{moderate} throughput} as \revision{several rows perform} a single addition. Furthermore, the relative area overhead can be high due to the intermediate cells~\cite{WallaceTree}, \revision{and the data-transfer between rows requires additional support for inter-row logic}.
    \item \emph{Bit-Serial Element-Parallel~\cite{Nishil, Ameer, SIMPLER, ComputeDRAM, SIMDRAM, FELIX, FloatPIM, NeuralCache, RACER, WangPatent, CRAM}:} This approach performs the operations of the bit-serial element-serial approach in parallel across all rows -- \emph{with the exact same latency} -- by exploiting column parallelism. That provides \emph{high throughput} as \revision{$r$ adders per array are performed simultaneously with identical latency}; however, the latency remains rather high as gates are performed serially (a single gate \revision{per cycle per $N$-bit adder}).
    \item \emph{Bit-Parallel Element-Parallel (parallel single-row)~\cite{RIME, MultPIM}:} This recent approach gains both \emph{higher throughput} and \emph{low latency} by introducing \emph{partitions}~\cite{FELIX, PartitionPIM} (see Section~\ref{sec:parallelFixed}). The partitions dynamically divide the rows to enable multiple concurrent column operations. The adder is still performed within a single row (and in parallel across all rows), yet multiple gates are performed within each row concurrently. The potential drawback is that partitions may introduce additional physical overhead; however, a recent work has proposed a low-overhead design~\cite{PartitionPIM}.
\end{enumerate}

We focus on the \emph{element-parallel} approaches as PIM is best suited for data-intensive applications which require high throughput. Figure~\ref{fig:Approaches}(b) summarizes our contributions for 16 variants of arithmetic functions, establishing a state-of-the-art foundation for arithmetic in PIM. We propose the first known general-purpose digital PIM algorithm for a \emph{majority} of the combinations (including cases previously considered impossible, such as floating-point addition~\cite{DRISA}), while also presenting minor and major ($>5\times$) improvements for others. We accomplish this via a combination of three methods: 
\begin{itemize}
    \item \emph{Variable Shift:} We develop a novel algorithm for element-parallel \emph{variable shifting:} each row $i$ starts with numbers $x^i$ and $t^i$, and the output is $x^i \ll t^i$. While this was previously considered \emph{impossible} (as each row can have a different shift~\cite{DRISA}), we attain this efficiently for the first time due to the combination of in-memory multiplexers and a logarithmic-shifter approach (without any custom periphery). We then tackle \emph{variable normalization}: each row $i$ starts with number $x^i$, and the output is $x^i$ left-shifted until the MSB is one. This is far more difficult as the shift-amount is unknown, and yet we attain this with latency nearly identical to variable shift due to a technique inspired by a binary search.
    \item \emph{Partition Toolbox:} We exploit a unique algorithmic topology enabled by partitions~\cite{PartitionPIM} towards an efficient \emph{toolbox} of general-purpose routines. These include both generalizations of routines proposed in MultPIM~\cite{MultPIM}, and two novel routines: \emph{reduction} -- reducing (e.g., AND, OR) bits of multiple partitions to a single bit, and \emph{prefixing} -- each partition receives the reduction of bits in partitions before it.
    \item \emph{Arithmetic Theory:} We provide a \emph{new perspective} on historical, lesser-known, algorithms in computer arithmetic, demonstrating their effectiveness with element-parallel in-memory computing for the first time. For example, we utilize Karatsuba~\cite{Karatsuba, Parhami} for bit-serial multiplication, parallel-prefix adders~\cite{BrentKung, Parhami} for bit-parallel addition, and carry-lookahead in division~\cite{IAD, ArithmeticAndLogic, Parhami} for bit-parallel division. Interestingly, some of these algorithms are not effective in traditional systems~\cite{Parhami, ArithmeticAndLogic}, yet unique considerations of PIM lead to their effectiveness here.
\end{itemize}

This paper is organized as follows. Section~\ref{sec:background} provides background on a wide variety of PIM technologies and their compatibility with the abstract model. We start in Section~\ref{sec:serialFixed} with bit-serial fixed-point arithmetic to establish the state-of-the-art approaches and our improvements, and continue in Section~\ref{sec:serialFloating} with bit-serial floating-point arithmetic. Section~\ref{sec:parallelFixed} then shifts to the bit-parallel fixed-point approach, introducing bit-parallel addition/subtraction/division for the first time and improving bit-parallel multiplication. We combine bit-serial floating-point and bit-parallel fixed-point in Section~\ref{sec:parallelFloating} to establish bit-parallel floating-point algorithms. Section~\ref{sec:results} evaluates AritPIM through a case study of memristive PIM implemented on a publicly-available cycle-accurate simulator, and Section~\ref{sec:conclusion} concludes this paper.

Throughout, we discuss abstract logic gates (e.g., AND, XOR, full-adder) and latency complexity (e.g., $O(N^2)$ where $N$ is the representation size) for generality and concise explanations, with Section~\ref{sec:results} reducing these to the underlying gates supported (e.g., NOR) and providing full implementations that prove correct results (e.g., matching the IEEE round-to-nearest ties-to-even \emph{exactly}). We refer to \emph{steps} rather than cycles, where each step performs a single abstract logic gate. Without loss of generality, as we focus on the element-parallel approaches, we often discuss gates performed within a \emph{single-row}, as the generalization to all rows and arrays is the trivial repetition of the gates. For fixed-point, we discuss unsigned numbers (for simplicity) yet the algorithms can extended to signed. Lastly, $v^i$ refers to the $i^{th}$ element of vector $v$, $x_i$ refers to bit $i$ of number $x$ (0 is the LSB), $x_{i:j}$ is bits $i$ (inclusive) through $j$ (exclusive), and $(x|y)$ concatenates $x$ (higher bits) and $y$ (lower bits).

\section{Digital Processing-in-Memory (PIM)}
\label{sec:background}

We discuss two examples of digital PIM, memristive stateful-logic~\cite{Nishil, FELIX, IMPLY, CRAM} and DRAM~\cite{Ambit, ComputeDRAM, SIMDRAM, DRISA}, as well as other PIM technologies. Overall, these include both architectures that are already commercially available, and emerging ones with vast potential that are backed by small experimental demonstrations. We show that \emph{all} are compatible with the abstract model assumed in this paper.

\subsection{Memristive Stateful-Logic}
\label{sec:background:memristive}

Memristive PIM with \emph{stateful-logic}~\cite{StatefulLogicReview} utilizes an emerging physical device, the memristor, which inherently supports both storage and logic at the exact same place. Large-scale memristive memories (storage only) with high density are already commercially available (e.g., Intel Optane), and several studies~\cite{IMPLY, StatefulLogicReview, LogicComputing} have experimentally demonstrated logic with memristors on a small-scale. Therefore, memristive digital PIM has the potential for an efficient large-scale practical implementation in the future. 

Memristors are two-terminal resistive devices with a unique property: their resistance may be modified through an applied voltage. By dividing the range of possible resistance values to a binary classification (i.e., low resistance corresponds to logical one, high resistance corresponds to logical zero), a memristor can store a single bit through its resistance. The value of a memristor is read by applying a low voltage and measuring the current to derive the resistance, and their value may be written by applying a high voltage. Stateful-logic performs logical gates~\cite{IMPLY, Nishil, FELIX, StatefulLogicReview}, between the resistance states of memristors: for memristors arranged in the circuit shown in Figure~\ref{fig:PIM}(a), the final resistance state of the bottom memristor (output) depends on the resistance states of the top memristors (input) at the beginning of the operation due to the voltage-divider structure which is formed. An additional initialization cycle is required for the output prior to the gate operation~\cite{Nishil}.

Notably, memristive PIM may also be implemented through MRAM devices~\cite{CRAM, 8697377}, and possesses several favorable characteristics. Rather than dividing the resistance spectrum into ranges for each logical value that may be prone to errors due to noise, the MRAM device inherently has two stable states that correspond to low and high resistance. Furthermore, MRAM devices possess excellent endurance compared to other memristive technologies. MRAM PIM~\cite{CRAM, 8697377} supports the same abstract model as Figure~1(a), and previous works~\cite{CRAM, 8697377, 8416761, WangPatent} have primarily explored the extension of MRAM computing to bit-parallel element-serial arithmetic.

Memristors are often arranged in dense crossbar-array structures. Such crossbars consist of vertical bitlines, horizontal wordlines, and memristors at the junctions. When considering the memristor as a binary storage element, a crossbar array essentially stores a binary matrix of information. The logic functionality of memristors is also compatible with the crossbar-array structure, as the same circuit observed in Figure~\ref{fig:PIM}(a) appears within a row of Figure~\ref{fig:PIM}(b). Interestingly, by applying voltages on bitlines, all of the rows perform the logic gate in parallel~\cite{Nishil}. Essentially, this leads to a logic operation on columns of bits in a single cycle; e.g., the bit-wise NOR of two columns computed and stored in a third column, in a single cycle~\cite{Nishil}. Therefore, the logic functionality within memristor crossbar arrays attains the abstract model presented in Figure~\ref{fig:PIM}(e) (see Section~\ref{sec:introduction}). 

\subsection{In-DRAM Logic}
\label{sec:background:DRAM}

Recent works have demonstrated that Dynamic Random Access Memory (DRAM) technology (the leading form of computer memory today) can also support in-memory bitwise logic computation by exploiting existing properties of DRAM cells~\cite{Ambit, DRISA, ComputeDRAM, SIMDRAM}. Various works have proposed minor changes (e.g., modifying decoders) to commercially-available DRAM memory in order to support in-memory logic gates~\cite{Ambit, DRISA, SIMDRAM}, while other works utilized commercially-available DRAM \emph{without modification} and already have large-scale experimental demonstrations~\cite{ComputeDRAM}. Therefore, in-DRAM logic has the potential for practical large-scale PIM implementation in the immediate future. 

DRAM memory utilizes capacitors to store information through the stored charge. The range of possible charge levels in a single capacitor is divided into a binary classification, thereby storing a single bit per capacitor. To minimize the charge leakage from the capacitor, a single DRAM cell also includes a transistor connected in series. The state of a DRAM cell is read by discharging the capacitor and measuring the voltage using a sense amplifier, while the state is written by applying a voltage on the capacitor (thereby changing the stored charge). When three DRAM cells are connected as illustrated in Figure~\ref{fig:PIM}(c), then all three cells stabilize on a state which is majority of the states prior to the operation~\cite{Ambit, SIMDRAM}. Therefore, $\Maj(A, B, C) = AB + AC + BC$ is inherently supported by the DRAM cells. By setting one of the three cells to a known value before performing the gate, the OR and AND gates can be performed~\cite{Ambit}. Inversion is supported with low-overhead modifications to sense amplifiers~\cite{Ambit}. Note that the true inputs are typically first copied to these cells (from other cells using copy operations) due to the destructive-input property of the logic gates.

DRAM cells are often arranged in dense sub-array structures. Such structures consist of a grid of DRAM cells, as shown in Figure~\ref{fig:PIM}(d), with buffers and sense amplifiers to the left\footnote{Without loss of generality, we consider the sub-array transposed for terminology identical to memristive stateful-logic.} of the array. We find that the circuit from Figure~\ref{fig:PIM}(c) also exists within every row of the sub-array in Figure~\ref{fig:PIM}(d), thereby enabling parallel bitwise logic-gate execution. Essentially, bit-wise operations on columns (e.g., majority, AND, OR, NOT) within sub-arrays can be performed in $O(1)$ cycles. Therefore, this attains the abstract model presented in Figure~\ref{fig:PIM}(e) (see Section~\ref{sec:introduction}).

\subsection{Additional PIM Technologies}
\label{sec:background:other}

This section briefly mentions additional PIM technologies and discusses their compatibility with the abstract model.

\subsubsection{SRAM} 
\label{sec:background:other:SRAM}

Computing via Static Random Access Memory (SRAM) technology performs logic operations using SRAM cells via the bitlines~\cite{ComputeCache, NeuralCache} (similar to in-DRAM computing). This technique enables bit-wise operations on columns with constant time within SRAM arrays -- supporting the abstract model. Further, a common form of in-SRAM processing involves associative computing~\cite{potter2012associative}, whereby the output is conditionally set according to a pattern search on the inputs. While such PIM architectures have  already been fabricated~\cite{ComputeSRAM, GSI}, the current limitation is that the memory size is limited by the low density of SRAM, thereby only supporting PIM for workloads with small datasets.

\subsubsection{Non-Stateful Memristive}
\label{sec:background:other:NonStatefulMemristive}

Memristive memory also supports non-stateful logic operations by performing the logic in the sense amplifiers of each crossbar (rather than using the memristors themselves). Pinatubo~\cite{Pinatubo} is a leading example of such an architecture, providing bit-wise parallel logic operations that adhere to the abstract model. The drawback of non-stateful logic is the higher latency and energy consumption over stateful-logic. 

\subsubsection{FeFET}
\label{sec:background:other:FeFET}

The Ferroelectric Field Effect Transistors (FeFET)~\cite{FeFET} technology is emerging as a form of memory which can also inherently support basic logic gates~\cite{ComputingInFeFET}. Similar to Pinatubo~\cite{Pinatubo}, the logic is performed within the sense amplifiers and with parallelism identical to the abstract model. FeFET has the potential for highly compact and energy-efficient implementation in the future~\cite{FeFET}.

\section{Bit-Serial Fixed-Point Arithmetic}
\label{sec:serialFixed}

This section details in-memory arithmetic algorithms for bit-serial computation on fixed-point numbers (the simplest case). We begin by introducing the state-of-the-art in-memory addition/subtraction algorithm~\cite{Nishil, MultPIM, CRAM, WangPatent} which is based on a ripple-carry approach. We continue by introducing the state-of-the-art in-memory multiplication algorithm~\cite{Ameer}, based on the shift-and-add approach, which we then improve through Karatsuba~\cite{Karatsuba, Parhami}. Finally, we address division by proposing a non-restoring algorithm which is optimized for element-parallel in-memory logic as it avoids the \emph{conditional} subtraction efficiently.

\subsection{Bit-Serial Fixed-Point Addition/Subtraction}
\label{sec:serialFixed:addSubtract}

We start with the simplest arithmetic functions, bit-serial fixed-point addition/subtraction, as an example of the format for in-memory arithmetic functions. These algorithms have been discussed in several previous PIM works~\cite{Nishil, CRAM, MultPIM, WangPatent} as classic examples to bit-serial arithmetic. We discuss only addition as subtraction can be derived from two's-complement addition. 

We begin with a formal description of the task. Assume that a single-row of an array contains two $N$-bit fixed-point numbers, $x$ and $y$, and some additional \emph{intermediate} cells that may be used freely. The algorithm may perform a single basic logic operation (e.g., NOR, full-adder) per \emph{step}, with the inputs and outputs of the gate being cells in the single-row. The state of the row at algorithm completion should contain $z=x+y$ stored in a pre-determined range of cells, as an $N+1$-bit fixed-point number (including the carry-bit). Note that the algorithm \emph{cannot read} the row at any time as this will not generalize to an element-parallel approach; rather, the algorithm must be based exclusively on \emph{data-flow}.

\begin{algorithm}[t]
    \small
    \centering
    \begin{algorithmic}[1]
    \renewcommand{\algorithmicrequire}{\textbf{Input:}}
    \renewcommand{\algorithmicensure}{\textbf{Output:}}
    \REQUIRE $N$-bit $x$, $N$-bit $y$ in a single row.
    \ENSURE $N+1$-bit result $z$ in the same row, where $z=x+y$.
    \STATE $carry \gets 0$
    \FOR{$i = 0, \hdots, N-1$}
    \STATEx \hspace{\algorithmicindent} \emph{Compute full-adder serially (e.g., serially executing 9 NOR gates and utilizing intermediate cells~\cite{Nishil, SIMPLER})}
    \STATE $z_i, carry \gets \FA(x_i, y_i, carry)$
    \ENDFOR
    \STATE $z_{N} \gets carry$
    
    \end{algorithmic}
    \caption{Bit-Serial Fixed-Point Addition}
    \label{alg:serialFixed:addition}
\end{algorithm}

Algorithm~\ref{alg:serialFixed:addition} details the ripple-carry approach to addition. The approach utilizes a single intermediate cell to store the current carry between the bits, and iteratively computes the full-adders from the LSB to the MSB. Notice that this differs from a traditional ripple-carry adder in that all gates (e.g., NOR) are applied serially to the rows of the memory (e.g., requiring cells in each row that store intermediate results). Overall, the latency is $O(N)$ steps. This general approach is optimized for different PIM technologies in recent works (e.g., CRAM~\cite{CRAM}, FELIX~\cite{FELIX} and MultPIM~\cite{MultPIM}). Such optimizations include efficient constructions of the full-adder from the logic gates supported by memristive PIM, and the storage of both the carry and the NOT carry throughout the iterations to save an additional cycle.

\subsection{Bit-Serial Fixed-Point Multiplication}
\label{sec:serialFixed:mult}

We begin this section by introducing the state-of-the-art approach introduced by Haj-Ali~\textit{et al.}~\cite{Ameer}, and continue by proposing an improvement based on the Karatsuba~\cite{Parhami, Karatsuba} recursion; while Karatsuba is typically only effective for extremely-wide numbers~\cite{Parhami} (e.g., thousands of digits), unique considerations for digital PIM enable improvement for regularly-sized numbers (e.g., $32$-bit). The overall task is: the row begins with $N$-bit fixed-point numbers $x$ and $y$, and contains the $2N$-bit result $z=x*y$ after completion.

\begin{algorithm}[t]
    \small
    \centering
    \begin{algorithmic}[1]
    \renewcommand{\algorithmicrequire}{\textbf{Input:}}
    \renewcommand{\algorithmicensure}{\textbf{Output:}}
    \REQUIRE $N$-bit $x$, $N$-bit $y$ in a single row.
    \ENSURE $2N$-bit result $z$ in the same row, where $z=x*y$.
    \STATEx \textit{Base case using previous~\cite{Ameer}. Note: $N_{thresh} \approx 20$.}
    \IF {$N \leq N_{thresh}$}
    \STATE $z \gets (0\cdots 0)_2$ \COMMENT{$2N$-bit.}
    \FOR{$i = 0, \hdots, N-1$}
    \STATEx \hspace{2em}\textit{Compute $p^i \gets \AND(x, y_i)$ (serially over $N$ bits).}
    \STATE \algorithmicfor\;{$j = 0, \hdots, N-1$}\;\algorithmicdo\;$p^i_j \gets \AND(x_j, y_i)$
    \STATEx \hspace{2em}\textit{Compute $z \gets z + (p^i \ll i)$ using Alg.~\ref{alg:serialFixed:addition}.}
    \STATE $z_{i:i+N+1} \gets z_{i:i+N} + p^i$
    \ENDFOR
    \STATEx \textit{Proposed Karatsuba recursion.}
    \ELSE
    \STATEx \hspace{\algorithmicindent}\textit{Compute using recursive calls and Alg.~\ref{alg:serialFixed:addition}.}\footnotemark
    \STATE $t_1' \gets (x_{0:N/2}+x_{N/2:N}) * (y_{0:N/2}+y_{N/2:N})$
    \STATE $t_0 \gets x_{0:N/2} * y_{0:N/2}, t_2 \gets x_{N/2:N} * y_{N/2:N}$
    \STATE $t_1 \gets t_1' - t_0 - t_2$
    \STATEx \hspace{\algorithmicindent}\textit{Compute $z \gets t_0 + t_1 \ll N/2 + t_2 \ll N$ using Alg.~\ref{alg:serialFixed:addition}.}
    \STATE $z \gets (t_2 | t_0)$
    \STATE $z_{N/2:2N} \gets z_{N/2:2N} + t_1$
    \ENDIF
    \end{algorithmic}
    \caption{Bit-Serial Fixed-Point Multiplication}
    \label{alg:serialFixed:multiplication}
\end{algorithm}
\footnotetext{By calculating $t_1'$ before $t_0$ and $t_2$, we can reuse the cells that stored $x_{0:N/2}+x_{N/2:N}$ and $y_{0:N/2}+y_{N/2:N}$ for storing $t_0$ and $t_2$ (thereby reducing the proposed algorithm's space complexity).}
The shift-and-add approach to multiplication constructs
\begin{equation}
x*y = \sum_{i=0}^{N-1} p^i \ll i, \quad \text{where }p^i = \AND(x, y_i).
\end{equation} 
This approach first initializes the $2N$-bit output $z$ to zero, and then iteratively adds $p^i \ll i$ to $z$ for each $i$. Only an $N$-bit adder is required due to the zeros contained in $p^i \ll i$ (only $N$ bits are non-zero) and $z$ (top $N-i$ bits are zero during the $i^{th}$ iteration). The base case in Algorithm~\ref{alg:serialFixed:multiplication} presents the state-of-the-art PIM algorithm~\cite{Ameer} which is based on this shift-and-add. The algorithm iterates over each $i$, computing the $i^{th}$ partial product, $p^i$, and adding it to the current sum $z$. The shift is not computed explicitly, rather the elements in $z$ are merely accessed shifted (the $N$ columns chosen from the $2N$ are different in each iteration); that is, the shift is \emph{simulated}. Overall, the latency is $O(N^2)$.

The Karatsuba~\cite{Karatsuba, Parhami} approach reduces asymptotic complexity to $O(N^{\log_23}) \approx O(N^{1.58})$ through an optimization to the recursive expression for $N$-bit multiplication. Consider $x=(x_1 | x_0),\ y = (y_1 | y_0)$, the separation of each $N$-bit number to the upper and lower bits ($N/2$-bit numbers). The naive recursion for multiplication notes that,

\begin{equation}
x * y = \underbrace{x_0y_0}_{t_0} + \underbrace{(x_0y_1 + x_1y_0)}_{t_1} \ll N/2 + \underbrace{x_1y_1}_{t_2} \ll N.
\label{eq:serialFixed:mult:recursion}
\end{equation}
Thus, in the naive recursion, an $N$-bit multiplication requires \emph{four} $N/2$-bit multiplications. The Karatsuba approach reduces the number of $N/2$-bit multiplications to \emph{three} by computing $t_1$ with a single $N/2$-bit multiplication,
\begin{equation}
t_1 = (x_1 + x_0)(y_1 + y_0) - t_2 - t_0.
\label{eq:serialFixed:mult:karatsuba}
\end{equation}
In traditional computing systems, this approach is only used for large-number multiplication~\cite{Parhami} as the objective is to minimize the critical path and bit-level access is not possible. Conversely, for \emph{bit-serial} in-memory computing, the latency is the total number of gates and arbitrary bit-level operations may be executed. Therefore, we utilize~(\ref{eq:serialFixed:mult:karatsuba}) with both the shift-and-add approach (base case) and Algorithm~\ref{alg:serialFixed:addition} to propose Algorithm~\ref{alg:serialFixed:multiplication}. The algorithm contains a base case of performing shift-and-add directly if $N$ is small, and otherwise performs the three recursive calls and computes the output according to~(\ref{eq:serialFixed:mult:recursion}) and~(\ref{eq:serialFixed:mult:karatsuba}). Overall, the latency is $O(N^{1.58})$, providing minor (yet significant) improvements starting\footnote{The crossover point $N_{thresh}$ depends on the latency for a recursive Karatsuba step (additions/subtractions and the smaller multiplications) compared to that of naive shift-and-add. The value \revision{$N_{thresh} \approx 20$} is found by increasing $N_{thresh}$ until Karatsuba reduces the overall latency. This crossover is \revision{largely} independent of the assumed logic gates (e.g., NOR) since the compared latencies (Karatsuba step and shift-and-add) are both \emph{primarily} proportional to the number of cycles per 1-bit full-adder; \revision{regardless, the exact value may vary slightly.}} at \revision{approximately $N \approx 20$}.

\subsection{Bit-Serial Fixed-Point Division}
\label{sec:serialFixed:div}
 
Here, we tackle the most complex operation out of the four elementary arithmetic operations: division. We begin with background on restoring and non-restoring division as theoretical concepts, and then continue by presenting a novel algorithm for in-memory division that is based on a customized non-restoring approach. While bit-serial division does exist~\cite{GraphLayout} (restoring), our proposed algorithm is based on a different approach (non-restoring) that is better suited for in-memory computing as it inherently avoids the \emph{conditional} subtraction. Such conditional operations (i.e., branches) are not directly compatible with the abstract model which requires that all rows operate in lockstep, thus they are converted to a sequence of operations that serially evaluate both branch outcomes and then select the output with a multiplexer. The overall task is defined as the integer\footnote{General fixed-point can be derived from such integer division.} division of $2N$-bit dividend $z$ by $N$-bit divisor $d$, with $N$-bit quotient $q$ and $N$-bit remainder $r$ (i.e., $z = qd + r$ for $r < d$).

Restoring division~\cite{Parhami} is based on \emph{conditionally} subtracting the divisor from the current remainder, as long as the result remains non-negative. Formally, $r^{j+1} = 2\cdot r^j - q_{j+1}\cdot d$, where $r^j$ is the $j^{th}$ partial remainder; restoring division chooses $q_{j+1} \in \{0,1\}$ such that $0 \leq r^{j+1} < d$, implying $q_{j+1}=1$ if $2r^j - d \geq 0$, else $q_{j+1}=0$. Thus, overall it iterates over: shifting the remainder (computing $2\cdot r^j$), and subtracting $d$ from $2\cdot r^j$ but \emph{only if} it remains non-negative. 

Non-restoring division~\cite{Parhami} enables intermediate negative remainders to avoid the conditional subtraction. Instead of conditional subtraction, it either subtracts or adds (the benefit is that the choice is known at the end of the \emph{previous} iteration). On a theoretical level, $q_i \in \{-1, +1\}$ rather than $q_i \in \{0, 1\}$. Practically, $-1, +1$ are stored as $0,1$, respectively, and corrections are performed at the end. Algorithm~\ref{alg:nonRestoringDiv} presents the \emph{theoretical} algorithm for binary non-restoring division, slightly modified with PIM in mind.

Algorithm~\ref{alg:serialFixed:division} details the proposed bit-serial fixed-point divider, based on the theoretical approach from Algorithm~\ref{alg:nonRestoringDiv}. Specifically, we utilize these optimizations:

\begin{algorithm}[t]
\small
 \caption{Non-Restoring Division \emph{(Theoretical)}}
 \begin{algorithmic}[1]
 \renewcommand{\algorithmicrequire}{\textbf{Input:}}
 \renewcommand{\algorithmicensure}{\textbf{Output:}}
 \REQUIRE $2N$-bit dividend $z$, $N$-bit divisor $d$
 \ENSURE $N$-bit quotient $q$, $N$-bit remainder $r$, where $z = qd + r$ and $r < d$
 
 \STATE $q \gets 2^{N-1}, r \gets z_{N-1:2N}$
 \label{alg:nonRestoringDiv:init}
 
 \FOR{$i = N-1, \hdots, 0$}
 \label{alg:nonRestoringDiv:for}
 
 \STATEx \hspace{\algorithmicindent}\textit{Add/subtract conditional on \textbf{previous} bit from $q$.} 
 
 \STATE \algorithmicif \ $q_i$ \algorithmicthen \ $r \gets r - d$ \algorithmicelse \ $r \gets r + d$
 \label{alg:nonRestoringDiv:for:if}
 
 \STATE $q_{i-1} \gets (r \geq 0)$
 \label{alg:nonRestoringDiv:for:qi}
 
 \STATE $r \gets 2\cdot r + z_{i-1}$
 \label{alg:nonRestoringDiv:for:Rshift}
 
 \ENDFOR
 \label{alg:nonRestoringDiv:endFor}
 
 \STATEx \textit{Non-restoring representation corrections.}
 
 \STATE $q \gets 2\cdot q + 1$
 \label{alg:nonRestoringDiv:Qfix}
 
 \STATE \algorithmicif \ $r < 0$ \algorithmicthen \ $q \gets q - 1, r \gets r + d$
 \label{alg:nonRestoringDiv:RfixIf}
 
 \end{algorithmic} 
 \label{alg:nonRestoringDiv}
 \end{algorithm}

\begin{enumerate}
    \item \emph{Conditional Addition/Subtraction:} Control-flow for conditional addition/subtraction (Alg.~\ref{alg:nonRestoringDiv}, Line~\ref{alg:nonRestoringDiv:for:if}) is replaced with data-flow. Similar to~\cite{GraphLayout}, we could compute $r-d$ and $r+d$ (serially), and implement a multiplexer to choose according to $q_i$; however, this results in large overhead for the multiplexer and computation of both $r-d$ and $r+d$. Instead, we utilize properties of the two's-complement representation by performing bit-wise exclusive-or of $d$ with $q_i$, and then performing addition between the result, $r$, and a carry-in of $q_i$ (Alg.~\ref{alg:serialFixed:division}, Line~\ref{alg:serialFixed:division:for:add})~\cite{Parhami}.\footnote{Note that this optimization is not possible with restoring division (e.g., with AND) as the sign of $r-d$ must be computed regardless.}
    \item \emph{Quotient-Bit Update:} Updating the quotient bit (Alg.~\ref{alg:nonRestoringDiv}, Line~\ref{alg:nonRestoringDiv:for:qi}) is achieved by checking the most-significant-bit of the remainder (two's complement). This is performed as part of the addition (Alg.~\ref{alg:serialFixed:division}, Line~\ref{alg:serialFixed:division:for:add}), without additional steps.
    \item \emph{Remainder Shifting:} Shifting the remainder (Alg.~\ref{alg:nonRestoringDiv}, Line~\ref{alg:nonRestoringDiv:for:Rshift}) is \emph{simulated} by the algorithm as the addition (Alg.~\ref{alg:serialFixed:division}, Line~\ref{alg:serialFixed:division:for:add}) already stores the shifted result.
    \item \emph{Non-Restoring Correction:} Correcting the representation mismatch (Alg.~\ref{alg:nonRestoringDiv}, Lines~\ref{alg:nonRestoringDiv:Qfix}, \ref{alg:nonRestoringDiv:RfixIf}) is converted to data-flow as follows. $2\cdot q$ is replaced by a shift of $q$ that is \emph{simulated} throughout previous references to $q_i$ (Alg.~\ref{alg:serialFixed:division}, Lines~\ref{alg:serialFixed:division:init}-\ref{alg:serialFixed:division:endFor}). The $+1$ (Alg.~\ref{alg:nonRestoringDiv}, Line~\ref{alg:nonRestoringDiv:Qfix}), and the subsequent conditional $-1$ (Alg.~\ref{alg:nonRestoringDiv}, Line~\ref{alg:nonRestoringDiv:RfixIf}) if $r<0$, are replaced with $q_0 \gets (r < 0)' = r_{N-1}'$ (Alg.~\ref{alg:serialFixed:division}, Line~\ref{alg:serialFixed:division:Qfix}). The conditional $r \gets r + d$ if $r < 0$ is replaced with the addition of the bitwise-and of $d$ and $r_{N-1}$ to $r$ (Alg.~\ref{alg:serialFixed:division}, Line~\ref{alg:serialFixed:division:Rfix}). 
\end{enumerate}
Overall, the latency of the proposed algorithm is $O(N^2)$ steps, with the constant significantly improved over the previous state-of-the-art restoring division~\cite{GraphLayout} primarily due to the non-restoring approach being better suited for PIM. Notice that a Karatsuba-style approach for division~\cite{KaratsubaDivision} does not improve the \emph{worst-case} complexity, and thus is not applicable to in-memory computing (as requiring purely data-flow implies that latency is limited by the worst case).

\begin{algorithm}[t]
\small
 \caption{Bit-Serial Fixed-Point Division}
 \begin{algorithmic}[1]
 \renewcommand{\algorithmicrequire}{\textbf{Input:}}
 \renewcommand{\algorithmicensure}{\textbf{Output:}}
    \REQUIRE $2N$-bit dividend $z$, $N$-bit divisor $d$ in a single row.
    \ENSURE $N$-bit quotient $q$, $N$-bit remainder $r$, in the same row, where $z=qd + r$ and $r < d$.
 
 \STATE $q_{N} \gets 1, r \gets z_{N-1:2N}$
 \label{alg:serialFixed:division:init}
 
 \FOR{$i = N-1, \hdots, 0$}
 \label{alg:serialFixed:division:for}
 
 \STATEx \hspace{\algorithmicindent}\textit{Compute using vectored-XOR (serially over the bits of $d$ with $q_{i+1}$) and Alg.~\ref{alg:serialFixed:addition}, with $q_{i+1}$ as carry-in.}

 \STATE $(q_i | r_{1:N}) \gets r + \XOR(d, q_{i+1}) + q_{i+1}$
 \label{alg:serialFixed:division:for:add}
 
 \STATE $r_0 \gets z_{i-1}$ 
 \label{alg:serialFixed:division:for:r0}
 
 \ENDFOR
 \label{alg:serialFixed:division:endFor}
 
 \STATE $q_0 \gets r_{N-1}'$ \COMMENT{Already shifted.}
 \label{alg:serialFixed:division:Qfix}
 
 \STATEx \textit{Compute using Alg.~\ref{alg:serialFixed:addition}.}
 
 \STATE $r \gets r + \AND(d, r_{N-1})$
 \label{alg:serialFixed:division:Rfix}
 
 \end{algorithmic} 
 \label{alg:serialFixed:division}
 \end{algorithm}

\section{Bit-Serial Floating-Point Arithmetic}
\label{sec:serialFloating}

This section expands the proposed bit-serial fixed-point arithmetic algorithms to floating-point numbers, thereby supporting the numerous applications that require floating-point accuracy. In-memory floating-point addition \emph{was previously considered impossible} as the shift operations in the alignment and normalization steps inherently require control-flow that is difficult to perform in an element-parallel manner~\cite{DRISA}. Conversely, we attain an in-memory unsigned floating-point addition algorithm \emph{for the first time} by first proposing a simple (but powerful and efficient) element-parallel variable shift routine. This routine receives numbers $x$ ($N_x$-bit) and $t$ ($N_t$-bit) in each array row, and outputs $x \gg t$ in that row (where the shift amount may be \emph{different for each row}). Furthermore, we generalize the proposed variable shifter to a variable normalization routine (left-shifts each number until the MSB is one) through a technique inspired by a \emph{binary search}, thereby also supporting \emph{signed} floating-point numbers. Lastly, we demonstrate floating-point multiplication/division by utilizing their fixed-point counterparts and the variable shift routine with $N_t=1$.

\subsection{Floating-Point Representation}
\label{sec:serialFloating:floatingBackground}

This section provides background on the floating-point representation~\cite{Parhami}. The representation is essentially inspired by the scientific number format, thereby representing a wide range of \emph{real} numbers (e.g., from $2^{-127}$ to $2^{127}$). Thus, the representation is crucial for large-scale applications that process real numbers. We consider the IEEE 754 format, which includes a sign-bit $s$ (1-bit), an unsigned exponent $e$ ($N_e$-bit), and an unsigned mantissa $m$ ($N_m$-bit). The value of the number $x$ is defined as
\begin{equation}
x = (-1)^{{s}} \cdot 2^{{e}-{b}} \cdot (1.{m}),
\end{equation}
where $b$ (bias) is a constant value (e.g., 127 for 32-bit), and the \say{$1.$} is known as the \emph{hidden-bit}. We adhere to the IEEE standard of \emph{round to nearest, ties to even}. For simplicity, we do not consider NaN/Inf/subnormals/overflows, yet the proposed algorithms could be modified to address these rare cases (if required by the application).

Arithmetic with floating-point numbers can be significantly more complex than fixed-point numbers due to the alignment and normalization steps, specifically with \emph{addition} and \emph{subtraction}. Counter-intuitively, multiplication and division algorithms for floating-point numbers are rather simple generalizations from fixed-point algorithms. Consider the multiplication of two floating-point numbers, $x_1$ and $x_2$,
\begin{multline}
     x_1 * x_2 =
     (-1)^{{s}_1} \cdot 2^{{e}_1-{b}} \cdot (1.{m}_1) \cdot (-1)^{{s}_2} \cdot 2^{{e}_2-{b}} \cdot (1.{m}_2) \\ =
     (-1)^{{s}_1 \oplus {s}_2} \cdot 2^{({e}_1+{e}_2-{b})-{b}} \cdot ((1.{m}_1) \cdot (1.{m}_2)).
\end{multline}
Thus, floating-point multiplication can be reduced to exclusive-or (XOR) for the sign-bit, addition for the exponent, and then fixed-point multiplication for the mantissas. Note that if the fixed-point multiplication results in a mantissa in the range $[2,4)$, then the mantissa is right-shifted once and the exponent is incremented~\cite{Parhami}. Floating-point division can be similarly reduced to the XOR of the sign-bits, the subtraction of the exponents, and then fixed-point division of the mantissas (with a single conditional left-shift and exponent decrement if the mantissa is in $[0.5, 1)$)~\cite{Parhami}.

Conversely, unsigned floating-point \emph{addition} is significantly more complex as the input numbers must be aligned to match exponents before the addition (e.g., $1.2 \cdot 10^8 + 4.0 \cdot 10^7 = 1.2 \cdot 10^8 + 0.40 \cdot 10^8 = 1.60 \cdot 10^8$ for base 10). Such alignment can lead to complex control-flow mechanisms. Furthermore, addition may also require an additional \emph{single} conditional right-shift normalization (e.g., $7.2 \cdot 10^2 + 4.1 \cdot 10^2 = 11.3 \cdot 10^2 = 1.13 \cdot 10^3$). Overall, the algorithm follows these steps: (1) computing the difference of the exponents, (2) shifting the mantissa of the number with the smaller exponent to match the number with the larger exponent, (3) adding the mantissas, and (4) normalizing the result.

Interestingly, signed addition, from which subtraction can be derived, requires a more complex normalization that left-shifts the mantissa several times until the MSB is one (e.g., $(1.0013 \cdot 10^1) + (-1.0000 \cdot 10^1) = 0.0013 \cdot 10^1 = 1.3 \cdot 10^{-2}$)~\cite{Parhami}. Therefore, steps (1), (2), (3), remain similar, while step (4) requires more complex mechanisms.

\subsection{Bit-Serial Variable Shift Routine}
\label{sec:serialFloating:variableShift}

\begin{algorithm}[t]
    \small
    \centering
    \begin{algorithmic}[1]
    \renewcommand{\algorithmicrequire}{\textbf{Input:}}
    \renewcommand{\algorithmicensure}{\textbf{Output:}}
    \REQUIRE $N_x$-bit $x$, $N_t$-bit $t$, in a single row.
    \ENSURE $N_x$-bit result $z$ in the same row, where $z = x \gg t$.
    
    \STATE $z \gets x$
    
    \FOR{$j = 0, \hdots, \min(N_t - 1, \log_2(N_x)-1)$}
    
    \STATEx \hspace{\algorithmicindent}\textit{Compute $z \gets \mux_{t_j}(z \gg 2^j, z)$ as follows:}
    
    \STATE \algorithmicfor\;{$i = 0, \hdots, N_x-2^j-1$}\;\algorithmicdo\;$z_i \gets \mux_{t_j}(z_{i+2^j}, z_i)$
    \label{alg:serialFloating:variableShift:firstInnerFor}
    
    \STATE \algorithmicfor\;{$i = N_x-2^j, \hdots N_x-1$}\;\algorithmicdo\;$z_i \gets \AND(\lnot\;t_j,z_i)$
    \label{alg:serialFloating:variableShift:secondInnerEnd}
    
    \ENDFOR
    
    \end{algorithmic}
    \caption{Bit-Serial Variable Shift Routine}
    \label{alg:serialFloating:variableShift}
\end{algorithm}

This section proposes the first in-memory algorithm for variable shifting through a simple-but-powerful approach that utilizes \emph{simulated} in-memory multiplexers and a logarithmic-shifter approach (requiring no custom periphery at all). The task is defined as follows: the row begins with $N_x$-bit \emph{integer} $x$ and $N_t$-bit \emph{integer} $t$, and the output is $z = x \gg t$\footnote{Left-shift is also supported due to symmetry.}. Note that the shift amount is variable (may be different in each row in the memory array), and recall that the algorithm must be based exclusively on data-flow.

We first address the special case of $N_t = 1$. In this case, $z = x \gg 1$ if $t_0=1$ and $z = x$ otherwise ($t_0=0$); that is, 
\begin{equation}
z = \mux_{t_0}(x \gg 1, x) = \mux_{t_0}(x_{1:N_x}, x_{0:N_x}).
\end{equation}
where $x_{1:N_x}$ is zero-extended with an additional bit (at the MSB). We note that a $2:1$ $N_x$-bit multiplexer can be derived from basic logic gates (e.g., NOR), thereby enabling the implementation of an \emph{in-memory multiplexer} algorithm~\cite{GraphLayout} (not a dedicated multiplexer circuit, rather a bit-serial algorithm derived from a sequence of gates). Therefore, we find that the case of $N_t = 1$ can be addressed with a single $2:1$ $N_x$-bit multiplexer algorithm, exclusively through data-flow.

\begin{figure}[!t]
    \centering
    \includegraphics[width=\linewidth]{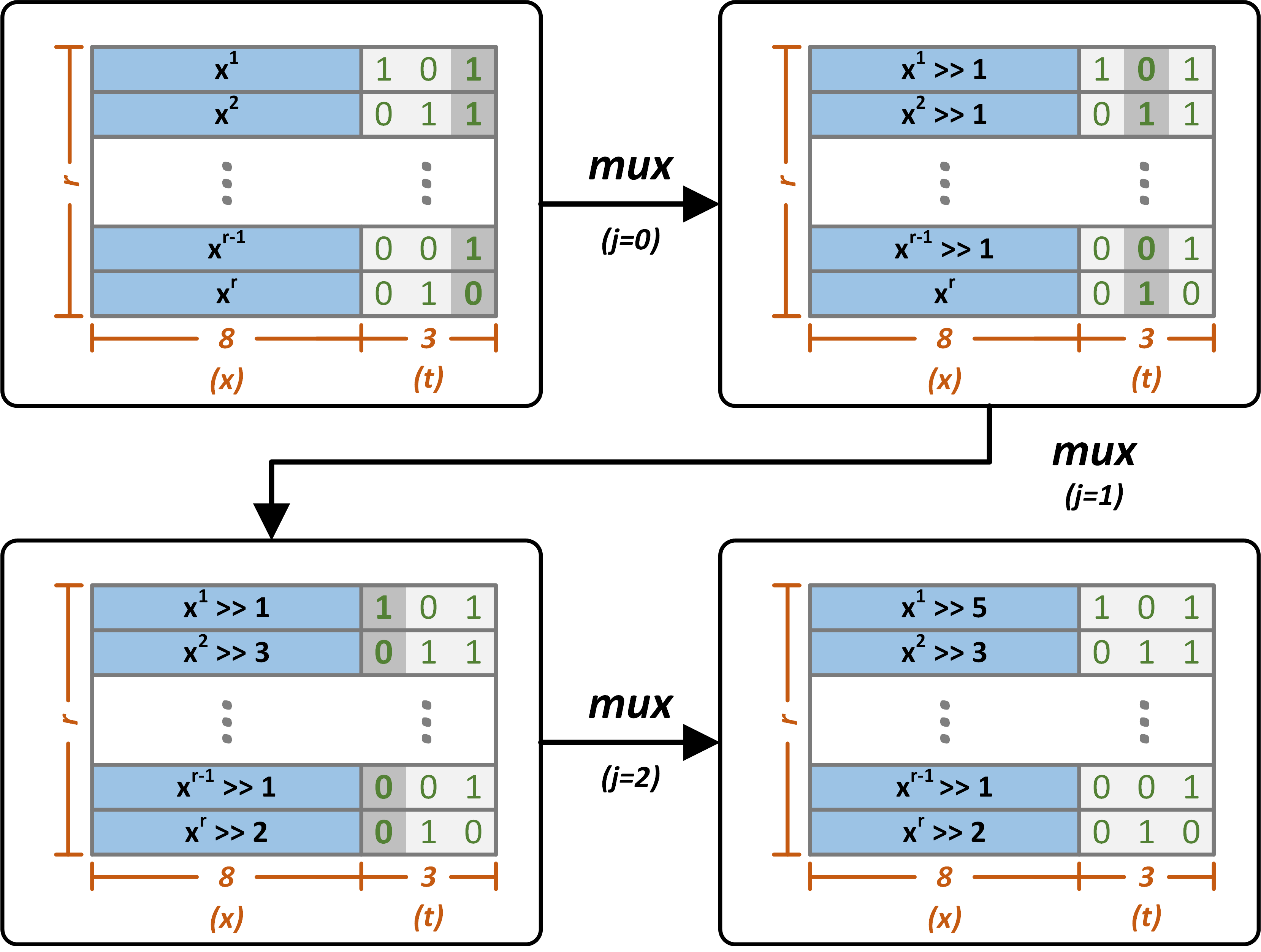}
    \caption{An example execution of the proposed in-memory bit-serial element-parallel \emph{variable shift} algorithm with $N_x=8$ and $N_t=3$. Each row $i$ right-shifts $x^i$ by its corresponding $t^i$ through $\log(N_x)$ iterations that construct the shift via multiplexers from shifts of size $2^0, 2^1, \cdots$.}
    \label{fig:serialFloating:variableShift}
    \vspace{-5pt}
\end{figure}

We efficiently extend the special case of $N_t=1$ to any $N_t$ through a \emph{logarithmic-shifter} approach. The basic concept of a logarithmic-shifter can be seen in the following example: shifting $x$ by 11 is identical to shifting $x$ by 1, then by $2$, and then by $8$ (the binary representation of 11). Therefore, the proposed algorithm begins with $z=x$ and then performs $\log_2(N_x)$ iterations\footnote{$\lceil\log(N_x)\rceil$ is computed as part of the algorithm compilation (i.e., by the controller as $N_x$ is constant).} where the $j^{th}$ iteration executes:
\begin{equation}
z = \mux_{t_j}(z \gg 2^j, z) = \mux_{t_j}(z_{2^j:N_x}, z_{0:N_x}).
\end{equation}
Algorithm~\ref{alg:serialFloating:variableShift} details the overall variable shift algorithm, and Figure~\ref{fig:serialFloating:variableShift} illustrates an example execution across all rows in an array (in parallel). The overall latency is $O(N_x\log (N_x))$ steps. Note that the zero-extension of $z_{2^j:N_x}$ for the upper $2^j$ bits is replaced by performing an AND gate rather than $2:1$ $1$-bit multiplexer at those indices.

\subsection{Bit-Serial Floating-Point Unsigned Addition}
\label{sec:serialFloating:additionUnsigned}
This section utilizes the novel variable-shift algorithm to propose the first in-memory floating-point addition algorithm for unsigned\footnote{This algorithm is only applicable when both numbers possess the same sign. See Section~\ref{sec:serialFloating:additionSigned} for an extension to signed addition.} numbers. Algorithm~\ref{alg:serialFloating:additionUnsigned} details the overall algorithm, essentially performing the theoretical steps for floating-point addition (see Section~\ref{sec:serialFloating:floatingBackground}), while utilizing the variable-shift routine (Algorithm~\ref{alg:serialFloating:variableShift}) for the alignment and single-bit normalization. Absolute value ($|\Delta e|$) is implemented by computing the exclusive-or of $\Delta e$ and $\Delta e \geq 0$ (MSB), and then adding $\Delta e \geq 0$ using Algorithm~\ref{alg:serialFixed:addition}. The \emph{hidden-bit} is addressed by concatenating a 1 to the mantissas and IEEE \emph{rounding} is addressed using sticky/round/guard bits~\cite{Parhami} (not shown); details are available in the code repository. We utilize an in-memory conditional-swap (in-memory multiplexers) to guarantee that the exponent of $x'$ is at least that of $y'$. The overall latency is $O(N_m\log(N_m) + N_e)$ steps.

\begin{algorithm}[t]
    \small
    \centering
    \begin{algorithmic}[1]
    \renewcommand{\algorithmicrequire}{\textbf{Input:}}
    \renewcommand{\algorithmicensure}{\textbf{Output:}}
    \REQUIRE Unsigned floating-point $x$, $y$ ($N_e$-bit exponents $x.e, y.e$, $N_m$-bit mantissas $x.m, y.m$) in a single row.
    \ENSURE Unsigned floating-point $z$ ($N_e$-bit exponent $z.e$, $N_m$-bit mantissa $z.m$) in the same row, where $z=x+y$.
    
    \STATEx \textit{Exponent difference using Alg.~\ref{alg:serialFixed:addition}:}
    
    \STATE $\Delta e \gets x.e - y.e$.
    \label{alg:serialFloating:additionUnsigned:exponentDiff}
    \STATE $z.e \gets \mux_{\Delta e \geq 0}(x.e, y.e)$ \COMMENT{Maximum.}
    
    \STATEx \textit{Conditional swap using vectored mux:}
    
    \STATE $x'.m \gets \mux_{\Delta e \geq 0}(x.m, y.m)$
    \STATE $y'.m \gets \mux_{\Delta e \geq 0}(y.m, x.m)$
    
    \STATEx \textit{Alignment using Alg.~\ref{alg:serialFloating:variableShift}:}
    
    \STATE $y'.m \gets y'.m \gg |{\Delta e}|$
    \label{alg:serialFloating:additionUnsigned:alignment}
    
    \STATEx \textit{Integer addition using Alg.~\ref{alg:serialFixed:addition}:}
    
    \STATE $z.m, carry \gets x'.m + y'.m$
    \label{alg:serialFloating:additionUnsigned:mantissaAddition}
    
    \STATEx \textit{Normalization using Alg.~\ref{alg:serialFloating:variableShift} and Alg.~\ref{alg:serialFixed:addition} (with $N_t=1$):}
    
    \STATE $z.m \gets z.m \gg carry$
    
    \STATE $z.e \gets z.e + carry$
    
    \end{algorithmic}
    \caption{Bit-Serial Floating-Point Addition (Unsigned)}
    \label{alg:serialFloating:additionUnsigned}
\end{algorithm}

\subsection{Bit-Serial Variable Normalization Routine}
\label{sec:serialFloating:variableNormalization}
To support signed floating-point addition/subtraction as well, we require an additional routine that is significantly more complex than variable shift: the input is $x$ in each row, and the output is \emph{both} $x$ left-shifted until the MSB is one and the shift amount stored as a number in the same row. The difficulty arises from the fact that the shift amount is not known in advance (and still may be different in each row). Naive attempts to compute the shift amount and then perform the variable shift routine (e.g., by using full-adders to count the number of leading zeros) will lead to a massive overhead in latency (approximately $300\%$ more steps). Conversely, inspired by a binary search, we propose a minor modification to the variable shift routine that adds a small number of steps (approximately $7\%$) and solves the variable normalization task exactly.

We begin by describing the approach through an example. Consider $N_x=8$, with $x=(00000110)$; thus, the \emph{desired output} is $z=(11000000)$ and $t=5=(101)$. The algorithm proceeds as follows. As the highest $N_x/2=4$ bits of $x$ are all zero (equivalently, their OR is zero), then $t_2=1$ (that is, $t \geq 4$) and thus we already set $z\gets x \ll 4 = (01100000)$. As the highest $N_x/4 = 2$ bits of $z$ are \emph{not} all zero (equivalently, their OR is non-zero), then $t_1 = 0$ and $z$ is not changed. As the highest $N_x/8 = 1$ bits of $z$ are all zero, then $t_0 = 1$ and $z \gets z \ll 1 = (11000000)$. Therefore, the algorithm correctly returned $z=(11000000)$ and $t=(101) = 5$. In general, for each $j=\log_2(N)-1, \hdots, 0$, the algorithm performs:
\begin{equation}
    t_j \gets \lnot(z_{N_x-2^j} \lor \cdots \lor z_{N_x-1}), \quad z \gets \mux_{t_j}(z \ll 2^j, z)
\end{equation}

\begin{algorithm}[t]
    \small
    \centering
    \begin{algorithmic}[1]
    \renewcommand{\algorithmicrequire}{\textbf{Input:}}
    \renewcommand{\algorithmicensure}{\textbf{Output:}}
    \REQUIRE $N_x$-bit $x$ in a single row.
    \ENSURE $N_x$-bit $z$, $\log_2(N_x)$-bit $t$, such that the number of leading zeros in $x$ is $t$, and $z = x \ll t$.
    
    \STATE $z \gets x$
    
    \FOR{$j = \log_2(N_x)-1, \hdots, 0$}
    
    \STATEx \hspace{\algorithmicindent}\textit{Compute $t_j \gets \lnot(z_{N_x-2^j} \lor \cdots \lor z_{N_x-1})$.}
    
    \STATE $temp \gets 0$
    \FOR{$i = N_x-2^j, \hdots N_x-1$}
    \STATE $temp \gets temp \lor z_i$
    \ENDFOR
    \STATE $t_j \gets \lnot\; temp$
    
    \STATEx \hspace{\algorithmicindent}\textit{Compute $z \gets \mux_{t_j}(z \ll 2^j, z)$.}
    
    \STATE {Omitted as nearly identical to Lines~\ref{alg:serialFloating:variableShift:firstInnerFor},~\ref{alg:serialFloating:variableShift:secondInnerEnd} of Alg.~\ref{alg:serialFloating:variableShift}.}
    
    \ENDFOR
    
    \end{algorithmic}
    \caption{Bit-Serial Variable Normalization Routine}
    \label{alg:serialFloating:variableNormalization}
\end{algorithm}

\begin{figure}[!t]
    \centering
    \includegraphics[width=\linewidth]{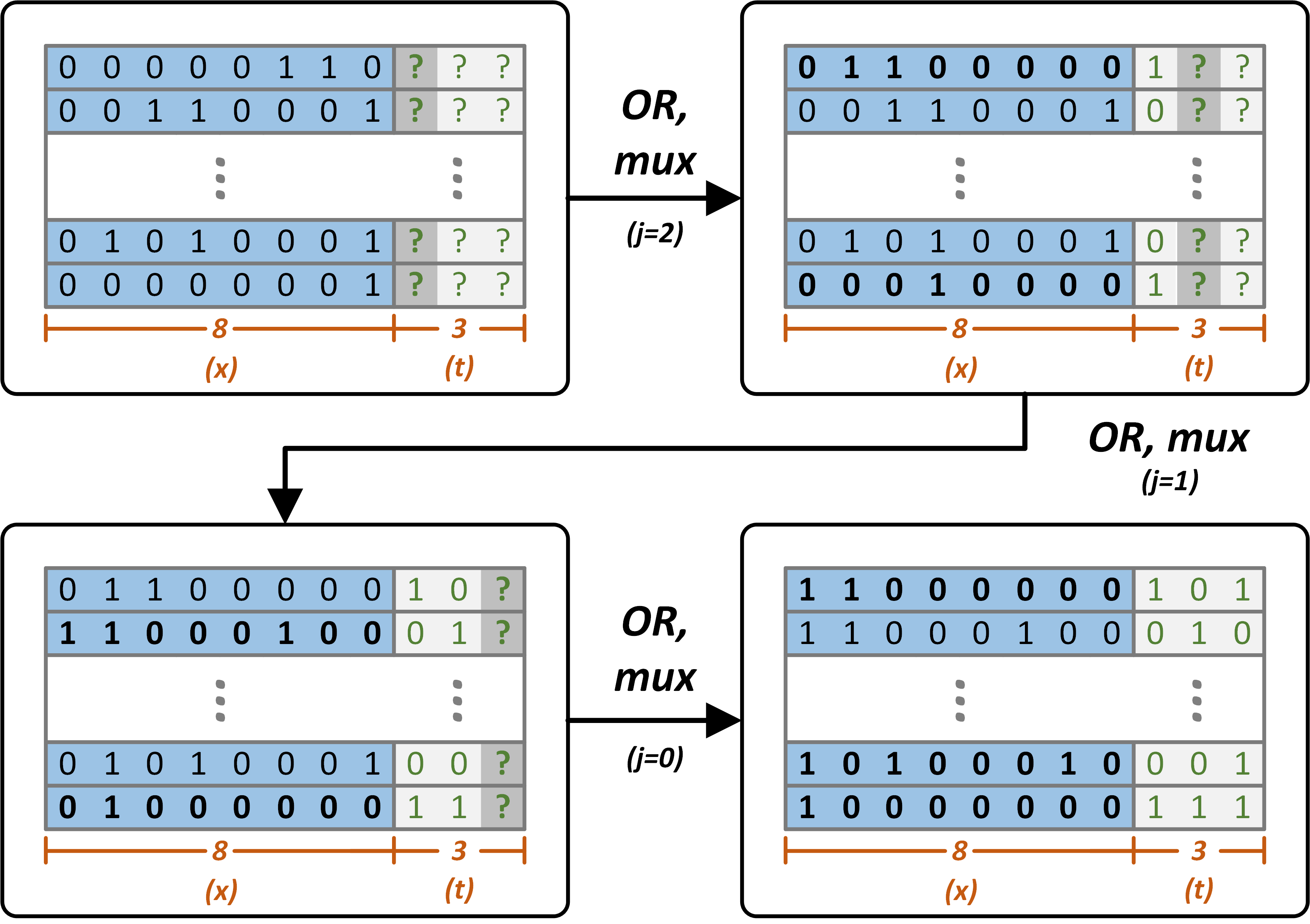}
    \caption{An example execution of the proposed in-memory bit-serial element-parallel \emph{variable normalization} algorithm with $N_x=8$. Each row $i$ left-shifts $x^i$ until the MSB is one, while also producing the shift amount $t^i$, via shift iterations that consist of a multiplexer and OR reduction.}
    \label{fig:serialFloating:variableNormalization}
\end{figure}
\noindent The above procedure is analogous to a binary search on the OR-prefix of $x$ (searching for the first 1 in the word): at each iteration, the size of the current search window (the interval which contains the first 1) is decreased two-fold, and the choice between the left/right intervals is performed by the in-memory multiplexer (always \say{pulling} the chosen interval to the left). Algorithm~\ref{alg:serialFloating:variableNormalization} details the overall proposed normalize-shift algorithm, which modifies the variable-shift algorithm by including the computation of $t_j$ within each iteration. Interestingly, this integrates the computation of $t$ \emph{within} the variable-shift iterations. The overall latency is
\begin{equation}
    O\left(\sum_{j=0}^{\log_2N_x-1}(N_x+2^j)\right) = O(N_x\log(N_x) + N_x);
\end{equation}
that is, the complexity remains identical to the $O(N_x \log(N_x))$ from variable-shift even though the task is significantly more difficult (since shift is unknown). Notice that the overhead over variable-shift is only the OR computations for $t$, which is $O(\sum_{j=0}^{\log_2N_x-1}2^j) = O(N_x)$ steps total, leading to the very low increase in latency over variable-shift (approximately $7\%$ for $N_x=24$). 

\subsection{Bit-Serial Floating-Point Signed Addition}
\label{sec:serialFloating:additionSigned}
This section utilizes the novel bit-serial variable normalization to propose the first in-memory \emph{signed} floating-point addition. Subtraction can be derived from such \emph{signed} addition by simply inverting the sign-bit of the second input. The proposed algorithm extends Algorithm~\ref{alg:serialFloating:additionUnsigned} as follows: $x'.m$ is replaced with $-x'.m$ if $x.s \neq y.s$~\cite{Parhami} (performed using data-flow through a technique similar to the Section~\ref{sec:serialFixed:div}), and Algorithm~\ref{alg:serialFloating:variableNormalization} is utilized at the end to left-normalize $z.m$ (in addition to the simpler single right-shift normalization using Algorithm~\ref{alg:serialFloating:variableShift}). The sign of the overall output is computed through data-flow via an expression involving $\Delta s = \XOR(x.s, y.s)$, whether $z.m$ was negative, and whether $x,y$ were swapped; details are available in the code repository. Overall, this algorithm possesses the same properties as the unsigned addition algorithm: the abstract PIM model is utilized without modifications (i.e., no custom periphery required), and the algorithm operates within a single-row via data-flow (enabling element-parallel execution). The overall latency remains $O(N_m\log(N_m) + N_e)$ steps, yet the constant increases slightly (over unsigned) as the variable-normalization (Algorithm~\ref{alg:serialFloating:variableNormalization}) is performed in addition to the variable-shifting (Algorithm~\ref{alg:serialFloating:variableShift}).

\subsection{Bit-Serial Floating-Point Multiplication/Division}
\label{sec:serialFloating:multDiv}
As detailed in Section~\ref{sec:serialFloating:floatingBackground}, these algorithms are derived naively from the corresponding bit-serial fixed-point algorithms to compute the result mantissa, and fixed-point addition/subtraction to compute the result exponent. The variable shift routine with $N_t=1$ is used for the final single-step normalization of the mantissa. The overall latency is $O({N_m}^{\log_2(3)} + N_e) \approx O({N_m}^{1.58} + N_e)$ steps for multiplication and $O({N_m}^2 + N_e)$ steps for division.

\subsection{Related In-Memory Floating-Point Works}
\label{sec:serialFloating:related}

The difficulty of in-memory element-parallel floating-point operations (e.g., variable shifting~\cite{DRISA}) has led to little research on the subject. For example, DRISA~\cite{DRISA} explicitly mentions the lack of a variable-shift routine obstructing the support for floating-point operations. Nonetheless, a few research works~\cite{FloatPIM, ReHy, ComputeSRAM} have previously attempted such floating-point operations. Yet, their algorithms require content-addressable-memory (CAM) functionality to be integrated within the arrays (no longer adhering to the abstract model). Specifically, they require the existence of a \emph{search} operation which can select certain rows for a mask based on the data stored in row, and then only apply column operations according to the mask. These search operations are exploited towards variable shift by iterating over all possible shift quantities (all values of $t$) and shifting only the rows which correspond to exactly that shift amount. 

Unfortunately, the integration of a CAM within memristive crossbar arrays may increase the memory area by approximately $12.5\times$~\cite{CAM}. This overhead has directly led to the algorithms \emph{not} being widely adopted~\cite{RACER, GraphLayout}; further, CAMs may not be compatible with other forms of PIM (e.g., DRAM). \emph{Conversely, our proposed bit-serial floating-point algorithms require no modifications, are compatible with many additional forms of PIM (e.g., DRAM), and, even without resorting to extra hardware, are faster than the previous works~\cite{FloatPIM, ReHy, ComputeSRAM} due to the logarithmic shifter and binary-search approaches.}

\section{Bit-Parallel Fixed-Point Arithmetic}
\label{sec:parallelFixed}

\begin{figure}[!t]
    \centering
    \includegraphics[width=\linewidth]{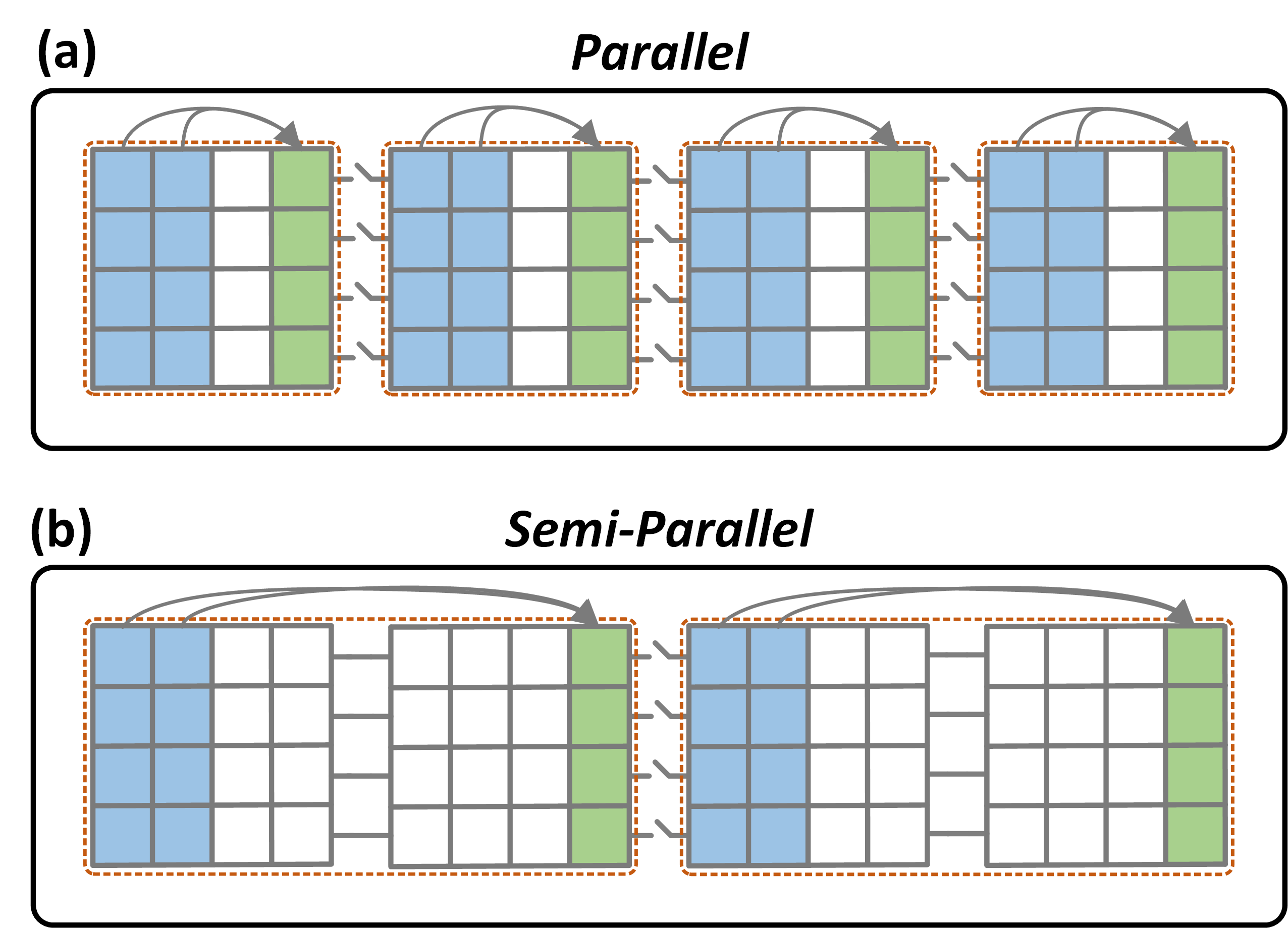}
    \caption{Partitions emerge when switches partition arrays to increase parallelism, for (a) fully-parallel (all switches disconnected) and (b) semi-parallel operations (some switches disconnected).}
    \label{fig:parallelFixed:partitionParallelism}
\end{figure}

We overcome an inherent limitation of the bit-serial element-parallel approaches by utilizing an emerging technique of \emph{partitions}~\cite{FELIX, PartitionPIM}, thereby introducing a highly-unique computation model that we exploit for bit-parallel element-parallel arithmetic. While the bit-serial element-parallel approach provides high throughput from the parallel computation across all rows and all arrays, the gates within each row are performed serially (e.g., one NOR at a time). This leads to high latency and low utilization of the potential computational power of PIM: all the $r\cdot c$ devices of each array have inherent logic capabilities, yet we only activate $O(r)$ devices per cycle. Partitions have recently emerged as a minor modification to PIM architectures that overcomes this limitation by enabling multiple concurrent column operations within a single array. Every array is horizontally divided into $k$ partitions, each sized $r \times (c/k)$, which are connected via $k-1$ sets of $r$ switches. This enables concurrent execution in different partitions when the switches are disconnected, and the ability to efficiently transfer data between partitions (using column operations) when some switches are connected through \emph{semi-parallel} operations, as illustrated in Figure~\ref{fig:parallelFixed:partitionParallelism}. We exploit this unique computational model towards a suite of \emph{bit-parallel element-parallel} fixed-point arithmetic algorithms that vastly improves latency. 

\begin{figure*}[!t]
    \centering
    \includegraphics[width=0.95\linewidth, trim={0cm, 0.3cm, 0cm, 0cm}]{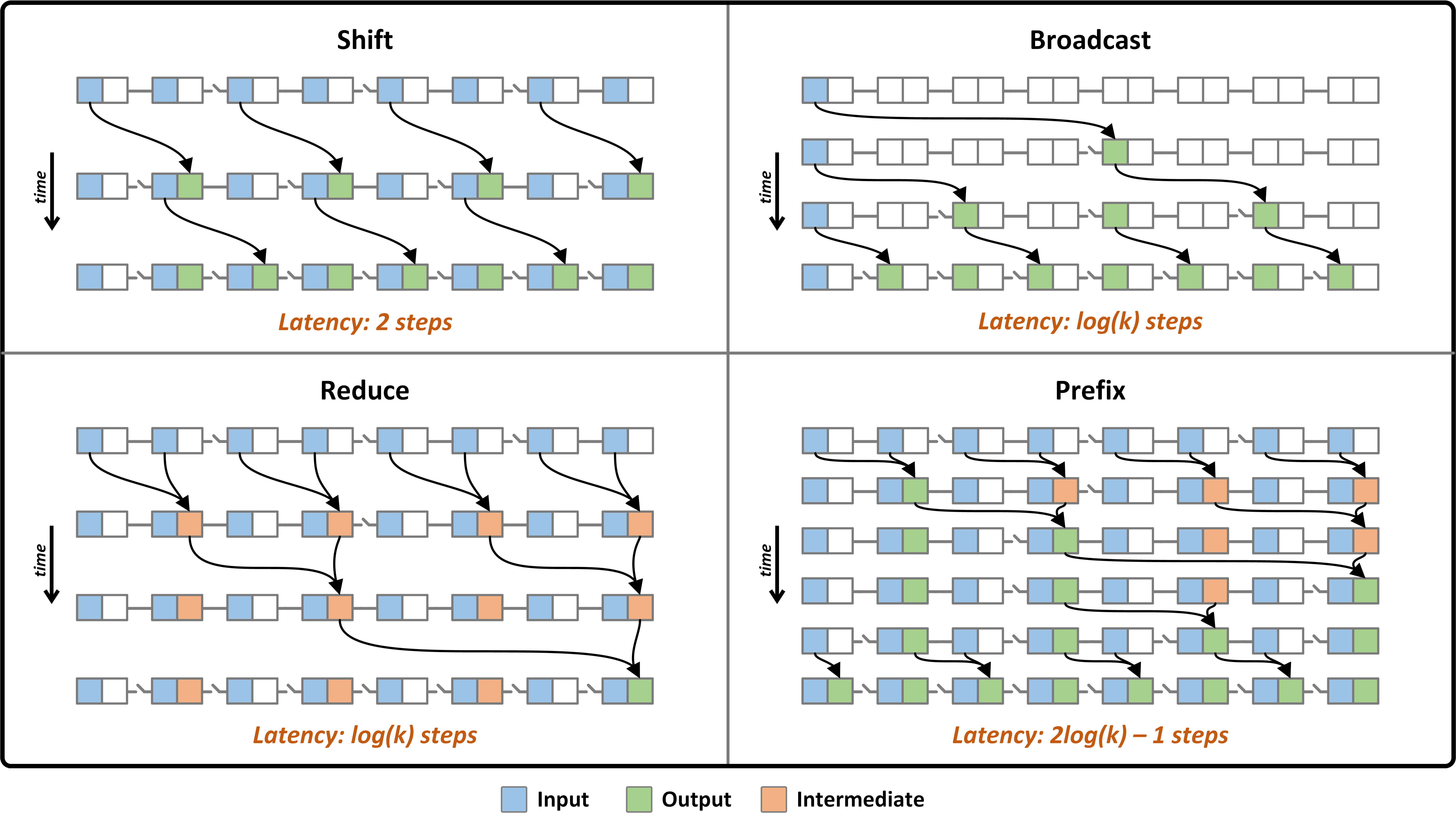}
    \caption{The proposed partition toolbox, extending techniques proposed in MultPIM~\cite{MultPIM}. The illustrations follow a row of partitions (each shown as two cells for simplicity) as they progress throughout the cycles (vertical axis represents time); the states of the switches (connected or disconnected) are illustrated in the connections between the partitions \revision{(the connectivity at time $t$ reflects the gates that occur between time $t$ and time $t+1$)}. The shift technique shifts a single bit between neighboring partitions using two steps, the broadcast technique broadcasts a single-bit from one partition to all partitions using $\log_2(k)$ steps, the proposed reduction technique reduces (e.g., AND) bits from all partitions to a single partition using $\log_2(k)$ steps, and the proposed prefix technique computes for each partition the reduction of the bits from partitions before it using $2\log_2(k)-1$ steps.}
    \label{fig:parallelFixed:partitionToolbox}
    \vspace{-5pt}
\end{figure*}

This section begins by providing further details on the computational model of partitions, and the support for partitions in PIM architectures. We then develop a general-purpose \emph{toolbox} for techniques that efficiently exploit the unique computational model of partitions. We first utilize this toolbox for fast fixed-point addition and subtraction through the parallel-prefix addition concept~\cite{BrentKung, Parhami}. We then present MultPIM~\cite{MultPIM}, the state-of-the-art for bit-parallel multiplication, and improve it by utilizing the proposed bit-parallel addition algorithm. Bit-parallel division is more complex than multiplication due to the conditional operations, and thus we utilize a lesser-known historical concept of carry-lookahead in division~\cite{IAD, Parhami, ArithmeticAndLogic} alongside a novel in-memory carry-lookahead routine.

\subsection{Partitions}
\label{sec:parallelFixed:partitions}

Partitions have recently emerged~\cite{FELIX, PartitionPIM} as a simple modification to PIM architectures that vastly improves parallelism with a highly-unique computational model. The dynamically-controlled switches dividing the partitions in each array allow merging adjacent partitions to perform gates amongst data stored in different partitions. At the extreme cases, we find that disconnecting all switches (parallel) provides maximal parallelism but with minimal flexibility (each gate is constrained to a single partition), and connecting all switches (serial) provides minimal parallelism but with maximal flexibility (each gate can access all of the columns of the array). Semi-parallelism refers to the case where only some switches are connected, providing intermediate parallelism and flexibility, see Figure~\ref{fig:parallelFixed:partitionParallelism}.

The implementation of partitions has been discussed in the context of memristive PIM using transistors~\cite{FELIX, PartitionPIM}, and should be applicable to DRAM PIM as well. Furthermore, partitions nearly attaining the model assumed in this paper are already included in a commercially-available SRAM processor~\cite{GSI}. Counter-intuitively, the overhead required for the switches is rather minuscule in comparison to the array size~\cite{FELIX, PartitionPIM}, with the more significant overhead being the periphery and control required to select all of the involved columns~\cite{PartitionPIM}. Therefore, PartitionPIM~\cite{PartitionPIM} proposes a reduced set of semi-parallel operations that requires various patterns to significantly reduce such overhead; in AritPIM, we assume this reduced set (minimal model~\cite{PartitionPIM}).

Partitions in arithmetic accelerate element-parallel algorithms by enabling multiple concurrent logic gates per function, see Figure~\ref{fig:Approaches}(a). We assume in this section that $k=N$ (representation size is exactly the number of partitions)\footnote{The proposed algorithms can be trivially extended to any $k > N$, and may also be generalized for the case of $k<N$.}. In this case, the inputs and outputs are stored in a \emph{strided} format: e.g., 32-bit integers are stored with a single bit per partition. This enables $O(1)$ latency for bitwise integer operations (e.g., integer OR of two 32-bit numbers) as all partitions operate in parallel. Arithmetic functions more complex than bitwise integer operations require semi-parallelism to share information between the different partitions. Only few works have explored bit-parallel arithmetic~\cite{RIME, MultPIM}.

\subsection{Partition Toolbox}
\label{sec:parallelFixed:toolbox}

We present a broad-range of partition techniques, constituting a general-purpose toolbox for bit-parallel algorithms. We start by briefly explaining and generalizing the \emph{broadcast} and \emph{shift} techniques from MultPIM~\cite{MultPIM}, and we then propose the novel \emph{reduction} and \emph{prefix} techniques. Prefix is complex as it is intuitively serial; yet, inspired by Brent-Kung~\cite{BrentKung}, we propose an efficient parallel algorithm with logarithmic time. For the remainder of the text, $k$ denotes the total number of partitions, $p_i$ denotes the $i^{th}$ partition, and $p_i.x$ refers to bit $x_i$ of number $x$ (which is stored in $p_i$). We also assume an abstract single-input \emph{identity} gate $\Id(A) = A$ for simplicity (e.g., implemented using two NOT gates).

\subsubsection{Shift}
\label{sec:parallelFixed:toolbox:shift}

Each partition stores a bit, and we shift the bits among the partitions. That is, $p_2$ gets the bit from $p_1$, $p_3$ gets the bit from $p_2$, etc. This is performed in two steps: odd partitions copy to even partitions, and then even partitions copy to odd~\cite{MultPIM} partitions, see Figure~\ref{fig:parallelFixed:partitionToolbox}. We generalize the above single-shift to shifting all bits $j$ partitions to the right (e.g, $p_{j+1}$ receives from $p_1$, $p_{j+2}$, receives from $p_2$, ...) in $j + 1$ steps with the $\ell^{th}$ step corresponding to all partitions $p_i$ such that $i = \ell \mod (j + 1)$. Notice that this is a deterministic shift (all rows shift by the same amount), unlike the variable shifting algorithm proposed in Section~\ref{sec:serialFloating}; Section~\ref{sec:parallelFloating} will combine these algorithms to attain bit-parallel variable shifting.

\subsubsection{Broadcast}
\label{sec:parallelFixed:toolbox:broadcast}

We desire to copy a single bit from a single partition (e.g., $p_1$) to all other partitions. This is achieved by first copying from $p_1$ to $p_{k/2+1}$, and then continuing recursively, in parallel, with the first half and the second half, as shown in Figure~\ref{fig:parallelFixed:partitionToolbox}. The parallel execution is achieved by disconnecting the switch between $p_{k/2}$ and $p_{k/2+1}$. In total, $\log_2(k)$ steps~\cite{MultPIM}.

\subsubsection{Reduction}
\label{sec:parallelFixed:reduction}

Each partition $p_i$ stores a bit $x_i$, and we output $x_1 \circ \cdots \circ x_k$, where $\circ$ is any associative operation (e.g., $AND, OR, XOR$). This is achieved via a logarithmic tree, as illustrated in Figure~\ref{fig:parallelFixed:partitionToolbox}. For example, in the first cycle, the partitions are split into adjacent pairs (e.g., $p_1$ and $p_2$ are connected, $p_3$ and $p_4$ are connected) and each section computes $\circ$ (storing the output in the partition with the larger index). The next cycle proceeds by connecting the partitions that contain the results in pairs (e.g., $p_2$ to $p_4$ are connected, $p_6$ to $p_8$ are connected). This continues following a logarithmic tree, until the last partition stores the overall reduction. Overall, $\log_2(k)$ steps in total.

\subsubsection{Prefix}
\label{sec:parallelFixed:prefix}
Each partition $p_i$ stores a bit $x_i$ as input, and each partition $p_i$ outputs $y_i = x_1 \circ \dots \circ x_i$, where $\circ$ is any associative operation (e.g., $AND, OR, XOR$). This generalizes the reduction technique as the last partition contains $x_1 \circ \cdots x_k$, and is far more complex as all partitions need to produce an output (the reduction of all bits up to that partition).

The naive algorithm will propagate through the partitions serially in $O(k)$ steps; that is, compute $y_2 = y_1 \circ x_2$ in $p_2$, then $y_3 = y_2 \circ x_3$ in $p_3$, and so forth. An improved algorithm relies on a recursive approach: first compute $y_{k/2}$ using the reduction technique in $O(\log_2 (k/2))$ steps, broadcast that result to the upper $k/2$ partitions (as all of their prefix expressions include the reduction of the lower half of partitions), and then proceed recursively. Yet, the recursive approach require $O({\log_2}^2(k))$ steps total. Conversely, we propose an algorithm inspired by Brent-Kung~\cite{BrentKung} that only requires $O(2\log_2(k)-1)$ steps and is essentially based on the unique combination of a single reduction operation followed by a single broadcast operation. Intuitively, the intermediate results computed in the reduction resemble a prefix at some indices, and then the broadcast \say{fills in the holes} for the other indices by causing internal propagation. An example execution is shown in Figure~\ref{fig:parallelFixed:partitionToolbox}, with the general algorithm provided in the code repository.

\subsection{Bit-Parallel Fixed-Point Addition/Subtraction}
\label{sec:parallelFixed:addition}

This section demonstrates the first bit-parallel addition algorithm, inspired by parallel-prefix carry-lookahead adders~\cite{BrentKung, Parhami} and utilizing the prefix technique from the partition toolbox. The task is defined as follows: each row begins with two $N$-bit fixed-point numbers, $x$ and $y$, stored in a strided format (i.e., each partition contains a single bit from $x$ and a single bit from $y$), and the algorithm computes $z=x+y$ and stores it also in a strided format.

The parallel-prefix carry-lookahead adder was developed as a carry-lookahead design that utilizes the unique prefix operation for low latency with efficient area and energy; the reader is referred to~\cite{Parhami} for a detailed explanation of the adder. Algorithm~\ref{alg:parallelFixed:addition} attains this prefix derivation using the prefix technique\footnote{Note that $\circ: \{0,1\}^2 \times \{0,1\}^2 \rightarrow \{0,1\}^2$ (each state is two bits). Therefore, the prefix technique is generalized to two bits per partition.} to propose the first in-memory bit-parallel addition algorithm. Overall, we require $O(\log(N))$ steps, improving the bit-serial state-of-the-art of $O(N)$ steps.

\begin{algorithm}[t]
    \small
    \centering
    \begin{algorithmic}[1]
    \renewcommand{\algorithmicrequire}{\textbf{Input:}}
    \renewcommand{\algorithmicensure}{\textbf{Output:}}
    \REQUIRE $N$-bit $x$, $N$-bit $y$ in a single row (strided format).
    \ENSURE $N$-bit result $z$ in the same row (strided format), where $z=x+y$.
    
    \STATEx \textit{Pre-computation of alive and generate bits:}
    
    \STATE $\forall i \;:\; p_i.A \gets \OR(p_i.x, p_i.y)$
  
    \STATE $\forall i \;:\; p_i.G \gets \AND(p_i.x, p_i.y)$
    
    \STATEx \textit{Prefix (using logarithmic prefix technique):}
    
    \STATE $\forall i \;:\; p_i.{GG}, p_i.{AA} \gets (p_i.G, p_i.A) \circ \cdots \circ (p_{0}.G, p_{0}.A)$ where $(g, a) \circ (\tilde{g}, \tilde{a}) = (g+a\tilde{g}, a\tilde{a})$.
    
    \STATEx \textit{Post-computation using shift and XOR:}
    
    \STATE $\forall i \;:\; p_{i+1}.c \gets p_i.GG$
    
    \STATE $\forall i \;:\; p_i.z \gets \XOR(p_i.x, p_i.y, p_i.c)$
    
    \end{algorithmic}
    \caption{Bit-Parallel Fixed-Point Addition}
    \label{alg:parallelFixed:addition}
\end{algorithm}

\subsection{Bit-Parallel Fixed-Point Multiplication}
\label{sec:parallelFixed:multiplication}

This section introduces MultPIM~\cite{MultPIM}, the state-of-the-art for bit-parallel multiplication which is based on the carry-save add-shift (CSAS) technique~\cite{FSP}, and then improves the algorithm via the proposed bit-parallel addition algorithm.

\begin{figure}[!t]
    \centering
    \includegraphics[width=\linewidth]{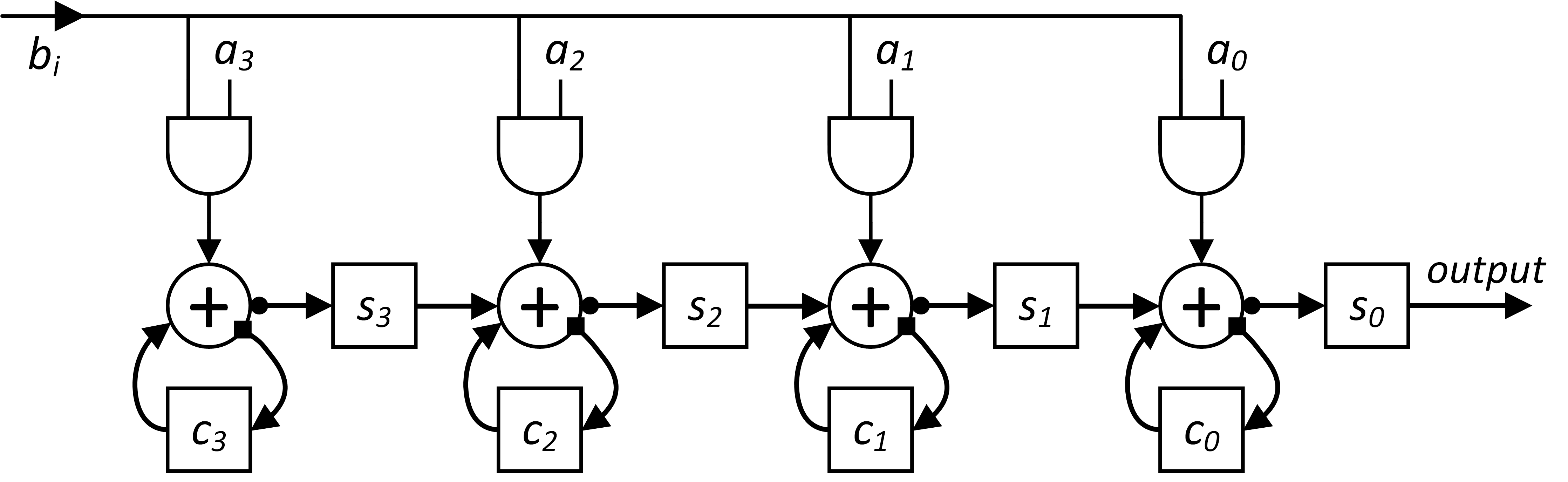}
    \caption{The carry-save add-shift (CSAS)~\cite{FSP, MultPIM} technique for multiplication. Circles are full-adders and squares are latches. Outputs of full-adders are marked as small circle (sum) and small square (carry).}
    \label{fig:parallelFixed:CSAS}
\end{figure}

The carry-save add-shift (CSAS) technique is a design for a latched multiplier circuit that computes the product of $N$-bit integers with $N$ parallel full-adder units. The motivation for CSAS begins with carry-save addition: a technique that avoids carry-propagation when computing the sum of many numbers. A carry-save adder receives three $N$-bit numbers, $X, Y, Z$, and outputs two $N$-bit numbers $S, C$, where $S + C = X + Y + Z$. This is done \emph{without carry-propagation} using $N$ full adders, where the $i^{th}$ adder receives $X_i, Y_i, Z_i$ and gives $S_i$ and $C_{i+1}$ (the $i^{th}$ carry-bit becomes the $i+1^{th}$ bit of $C$). To add many numbers, $X_1, \hdots, X_n$, we use carry-save adders to reduce it to a sum of only two numbers (fast as carry propagation is avoided), and then perform that using a single regular adder. Recall from Section~\ref{sec:serialFixed:mult} that multiplication can be expressed as the sum of many partial products; the fundamental idea of CSAS is using a carry-save adder for that addition. Similar to Section~\ref{sec:serialFixed:mult}, the circuit exploits the zeros in the partial products and in the running sum. Figure~\ref{fig:parallelFixed:CSAS} demonstrates the CSAS technique through a latched circuit design. For the $i^{th}$ iteration out of $N$, the bit $b_i$ is provided, all of the $N$ full-adders perform the carry-save addition between the $i^{th}$ partial product and the current running sum, and then all of the sum bits are shifted to the right and the last partition outputs $(a*b)_i$. These $N$ iterations compute the lower $N$ bits of $a*b$; for the upper $N$ bits, either a regular adder is used, or an additional $N$ stages feed zeros instead of $b_i$ for $N$ iterations.

\begin{algorithm}[t]
\small
 \caption{Bit-Parallel Fixed-Point Multiplication}
 \begin{algorithmic}[1]
 \renewcommand{\algorithmicrequire}{\textbf{Input:}}
 \renewcommand{\algorithmicensure}{\textbf{Output:}}
 \REQUIRE $N$-bit $x$, $N$-bit $y$ in a single row (strided format).
 \ENSURE $N$-bit results $z, w$ in the same row (strided format), where $(w|z)=x*y$.
  \STATE $\forall i \;:\; p_i.c, p_i.s \gets 0$
  \label{alg:fsm:SC}
  \FOR {$i = 0, \hdots, N-1$}
  \label{alg:fsm:firstFor}
  \STATEx \hspace{\algorithmicindent}\textit{Broadcast of $b_i$ to all partitions:}
  \STATE $\forall j \;:\; p_j.b' \gets p_i.b$
  \label{alg:fsm:firstFor:b}
  \STATEx \hspace{\algorithmicindent}\textit{Partial product computation:}
  \STATE $\forall j \;:\; p_j.ab \gets p_j.a \cdot p_j.b'$
  \label{alg:fsm:firstFor:ab}
  \STATEx \hspace{\algorithmicindent}\textit{Carry-save addition, with shift:}
  \STATE $\forall j \;:\; p_j.s, p_j.c \gets FA(p_j.s,p_j.c,p_j.ab)$
  \label{alg:fsm:firstFor:fa}
  \STATE $\forall j \;:\; p_{j}.s \gets p_{j+1}.s$
  \label{alg:fsm:firstFor:shift}
  \STATE $p_{i}.z \gets p_0.s$
  \ENDFOR
  \label{alg:fsm:firstFor:end}
  \STATEx \textit{Proposed Final Addition using Alg.~\ref{alg:parallelFixed:addition}.}
  \STATE $w \gets s + c$
 \end{algorithmic} 
 \label{alg:parallelFixed:mult}
 \end{algorithm}

MultPIM~\cite{MultPIM} utilizes the CSAS technique alongside the shift and broadcast techniques to propose an efficient bit-parallel multiplier. Each full-adder unit becomes a partition, thereby enabling the computation of all the full-adders in parallel. Furthermore, the partial products are also computed with parallelism as the broadcast technique is utilized to copy $b_i$ to all of the partitions in the $i^{th}$ iteration. Lastly, the shift technique is utilized to move the sum bits between the partitions efficiently. By performing $N$ such iterations, MultPIM computes the lower $N$ bits of the product. These steps are presented in Lines~\ref{alg:fsm:SC}-\ref{alg:fsm:firstFor:end} of Algorithm~\ref{alg:parallelFixed:mult}.\footnote{Note that MultPIM~\cite{MultPIM} was slightly modified to support strided format for both inputs and outputs.} To compute the upper $N$ bits of the product, MultPIM proceeds with an additional $N$ iterations where $0$ is provided instead of $b_i$ (full-adders are replaced with half-adders), instead of computing the sum of $S$ and $C$ directly. This choice was due to the best adder at the time requiring $O(N)$ cycles, and thus the latency was identical to performing an additional $N$ iterations of $O(1)$ cycles each. Yet, the proposed bit-parallel adder from Section~\ref{sec:parallelFixed:addition} can now be utilized instead, providing a significant advantage as $O(N)$ is reduced to $O(\log N)$ cycles and energy consumption (gate count) is reduced by approximately $1.6\times$. Algorithm~\ref{alg:parallelFixed:mult} provides the proposed algorithm for bit-parallel multiplication which utilizes this optimization. Overall, we require $O(N\log(N) + \log(N))$ steps, improving over the $O(N\log(N) + N)$ of MultPIM.
 
\subsection{Bit-Parallel Fixed-Point Division}
\label{sec:parallelFixed:division}

We present the first bit-parallel divider by combining the concepts of \emph{carry-save} and \emph{carry-lookahead}. Division is far more complex than multiplication due to the conditional subtraction. While the conditional subtraction was avoided in Section~\ref{sec:serialFixed:div} by replacing the control-flow with data-flow, the challenge here is that the sign of $R$ (determined from the MSB) is not known if $R$ is in carry-save format (represented as $R = S + C$). Therefore, this provides an inherent contradiction between the carry-save format and division. We overcome this by utilizing a carry-lookahead design similar to Section~\ref{sec:parallelFixed:addition} to efficiently compute only the sign of $S+C$.

\subsubsection{Carry-Save Carry-Lookahead (CSCL)}

This section proposes the carry-save carry-lookahead (CSCL) technique for latched division which combines the carry-save technique from CSAS with carry-lookahead to predict the sign of $R$. The technique is based on a lesser-known design for parallel division arrays~\cite{IAD} that is traditionally not used due to the complex layout~\cite{Parhami, ArithmeticAndLogic}.

The carry-save carry-lookahead (CSCL) division technique is shown in Figure~\ref{fig:CSCL}, based on CSAS and Algorithm~\ref{alg:nonRestoringDiv}. A single-bit of $z$ is inputted and a single-bit of $q$ is outputted at each of the $N$ iterations. Throughout the division, $r$ is represented by $s,c$ such that $r = s + c$ (carry-save format), and addition with $r$ (Alg.~\ref{alg:nonRestoringDiv}, Line~\ref{alg:nonRestoringDiv:for:if}) is achieved via carry-save addition. The sign bit of $r$ (Alg.~\ref{alg:nonRestoringDiv}, Line~\ref{alg:nonRestoringDiv:for:qi}) is computed from $s,c$ via carry-lookahead for the most-significant-bit. Shifting $r$ (Alg.~\ref{alg:nonRestoringDiv}, Line~\ref{alg:nonRestoringDiv:for:Rshift}) is achieved by shifting both $s$ and $c$. The main aspects of CSCL are:

\begin{figure}[!t]
    \centering
    \includegraphics[width=\linewidth]{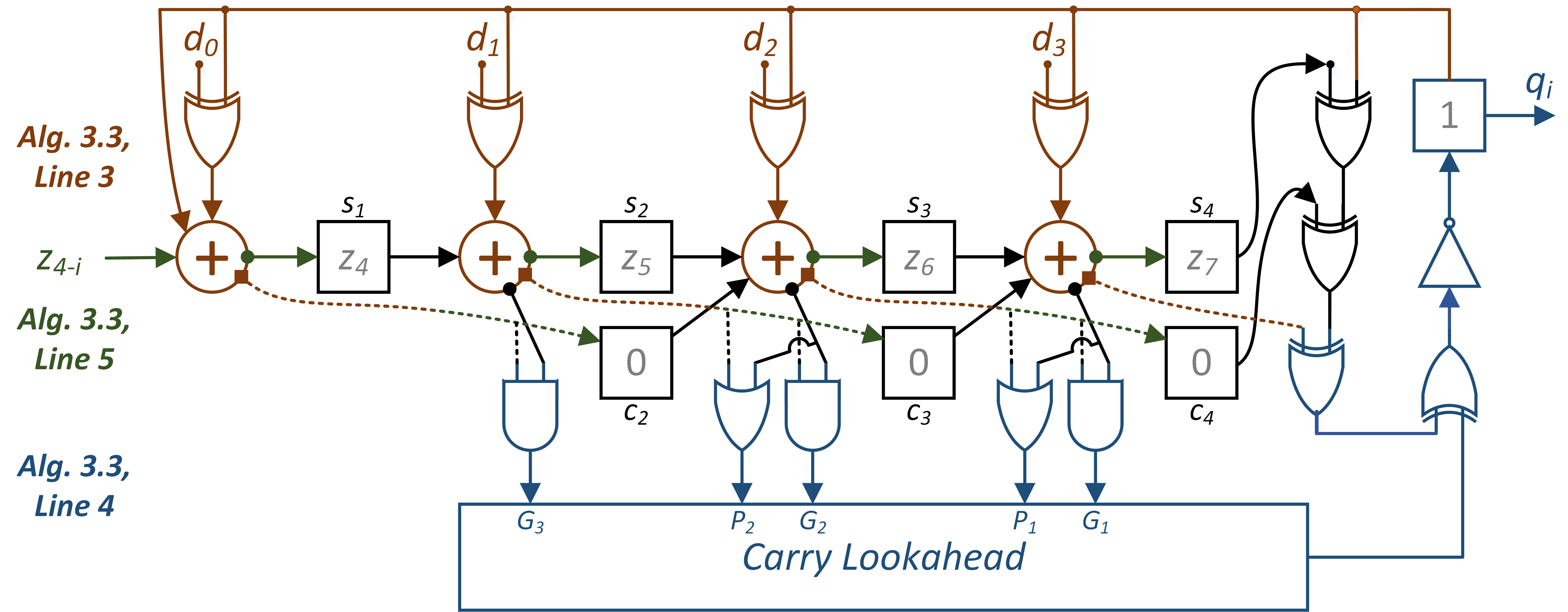}
    \caption{The latched carry-save carry-lookahead division circuit for $N=4$. Latches are squares and full-adders are circles. Solid/dashed line distinction indicates wires that do not intersect. Initial values are in gray.}
    \label{fig:CSCL}
\end{figure}

\begin{algorithm}[t]
\small
 \caption{Bit-Parallel Fixed-Point Division}
 \begin{algorithmic}[1]
 \renewcommand{\algorithmicrequire}{\textbf{Input:}}
 \renewcommand{\algorithmicensure}{\textbf{Output:}}
 \REQUIRE $2N$-bit dividend $(w|z)$, $N$-bit divisor $d$ in a single row (strided format).
 \ENSURE $N$-bit quotient $q$, $N$-bit remainder $r$ in the same row (strided format), where $(w|z) = qd + r$.
 
 \STATE $\forall i \;:\; p_i.s \gets p_i.w, p_i.c \gets 0$
 
 \FOR {$i = N-1, \hdots, 0$}

  \Statex \hspace{\algorithmicindent}\textit{Broadcast of $q_{i+1}$ to all partitions.}
  
  \STATE $\forall j \;:\; p_j.q' \gets p_{i+1}.q$
  
  \Statex \hspace{\algorithmicindent}\textit{Conditional addition/subtraction.}
  
  \STATE $\forall j \;:\; p_j.dq \gets \XOR(p_j.d, p_j.q')$
  
  \STATE $\forall j \;:\; p_j.s, p_j.c \gets \FA(p_j.s,p_j.c,p_j.dq)$
  
  \Statex \hspace{\algorithmicindent}\textit{First carry shift.}
  \STATE $\forall j \;:\; p_{j+1}.c \gets p_j.c$
  
  \Statex \hspace{\algorithmicindent}\textit{Compute $carry$ the carry of $s + c$:}
  
  \STATE Same as Alg.~\ref{alg:parallelFixed:addition} with prefix replaced with reduction.
  
  \STATE $p_i.q \gets \XNOR(p_{N-1}.s, p_{N-1}.c, p_{N}.s, p_{N}.c, carry)$
  \label{alg:parallelDivider:firstFor:qiUpdate}
  
  \Statex \hspace{\algorithmicindent}\textit{Remainder shifting:}
  
  \STATE $\forall j \;:\; p_{j+1}.s \gets p_j.s, p_{j+1}.c \gets p_j.c$
  
  \ENDFOR
  \Statex \textit{Final remainder computation using Alg.~\ref{alg:parallelFixed:addition}.}
  \STATE $r = s + c + \AND(d, \NOT(q_0))$
 \end{algorithmic} 
 \label{alg:parallelFixed:division}
 \end{algorithm}

\begin{itemize}
    \item \emph{Carry-Save Format (Figure~\ref{fig:CSCL}, black):} The remainder $r$ is represented via $s,c$ such that $r = s + c$, where $s,c$ are stored in latches $s_1, \hdots, s_N$ and $c_2, \hdots, c_N$.
    \item \emph{Carry-Save Addition (Figure~\ref{fig:CSCL}, orange):} Similar to the bit-serial divider, the conditional addition/subtraction (Alg.~\ref{alg:nonRestoringDiv}, Line~\ref{alg:nonRestoringDiv:for:if}) is achieved via XOR (Alg.~\ref{alg:serialFixed:division}, Line~\ref{alg:serialFixed:division:for:add}). The addition is performed via the carry-save technique, utilizing $N$ full-adders, and shifting the carry bits once (dashed orange arrows).
    \item \emph{Carry Lookahead (Figure~\ref{fig:CSCL}, blue):} The sign bit of $r$ (required in Alg.~\ref{alg:nonRestoringDiv}, Line~\ref{alg:nonRestoringDiv:for:qi}) is computed directly via carry-lookahead for the addition $s+c$.
    \item \emph{Remainder Shifting (Figure~\ref{fig:CSCL}, green):} The remainder shifting (required in Alg.~\ref{alg:nonRestoringDiv}, Line~\ref{alg:nonRestoringDiv:for:Rshift}) is achieved by shifting both sum and carry bits once to the right. 
\end{itemize}

\subsubsection{Proposed Algorithm}

The proposed CSCL technique is used in Algorithm~\ref{alg:parallelFixed:division}. As in bit-parallel multiplication, each full-adder is represented via a partition, and inter-partition communication (e.g., carry-save, remainder shifting) is achieved with the toolbox (e.g., shifting, broadcasting). The carry-lookahead for the last carry is performed by modifying Section~\ref{sec:parallelFixed:addition} to perform \emph{reduction} rather than \emph{prefix}. This computes the last carry in logarithmic time, and is faster than computing all carries as reduction is faster than prefix\footnote{This leads to the benefit of the proposed algorithm over merely implementing Algorithm~\ref{alg:nonRestoringDiv} with Algorithm~\ref{alg:parallelFixed:addition}.}. Overall, this is $O(N\log(N))$ steps, while the bit-serial state-of-the-art is $O(N^2)$. 

\section{Bit-Parallel Floating-Point Arithmetic}
\label{sec:parallelFloating}

We merge the ideas from Section~\ref{sec:serialFloating} (bit-serial floating-point) and Section~\ref{sec:parallelFixed} (bit-parallel fixed-point) for bit-parallel floating-point arithmetic. Recall that the algorithms from Section~\ref{sec:serialFloating} relied on the variable shifting/normalization and fixed-point counterparts from Section~\ref{sec:serialFixed}. We now show \emph{bit-parallel} variable shifting/normalization, and the extension to bit-parallel floating-point arithmetic follows by replacing the algorithms from Section~\ref{sec:serialFixed} with those from Section~\ref{sec:parallelFixed}. To allow fixed and floating point in the same crossbar, floating-point numbers are stored in strided format like fixed-point numbers (e.g., 32-bit floats are stored across 32 partitions). 


\begin{algorithm}[t]
    \small
    \centering
    \begin{algorithmic}[1]
    \renewcommand{\algorithmicrequire}{\textbf{Input:}}
    \renewcommand{\algorithmicensure}{\textbf{Output:}}
    \REQUIRE $N_x$-bit $x$, $N_t$-bit $t$, in a single row (strided format).
    \ENSURE $N_x$-bit $z = x \gg t$ in the same row (strided format).
    
    \STATE $\forall i \;:\; p_i.z \gets p_i.x$
    
    \FOR{$j = 0, \hdots, \min(N_t-1,\log_2(N_x)-1)$}
    
    \STATEx \hspace{\algorithmicindent}\textit{Compute $z \gets \mux_{t_j}(z \gg 2^j, z)$ as follows:}
    
    \STATEx \hspace{\algorithmicindent}\textit{Generalized shift technique:}
    \STATE $\forall i \;:\; p_i.z' \gets p_{i+2^j}.z$
    \STATEx \hspace{\algorithmicindent}\textit{Broadcast technique:}
    \STATE $\forall i \::\; p_i.s \gets p_j.t$
    \STATEx \hspace{\algorithmicindent}\textit{Parallel multiplexer:}
    \STATE $\forall i \;:\; p_i.z \gets \mux_{p_i.s}(p_i.z', p_i.z)$
    
    \ENDFOR
    
    \end{algorithmic}
    \caption{Bit-Parallel Variable Shift Routine}
    \label{alg:parallelFloating:variableShift}
    
\end{algorithm}

We extend the bit-serial variable-shift and variable-normalization routines from Section~\ref{sec:serialFloating:variableShift} using the partition toolbox. 
For bit-parallel \emph{variable-shift}, we utilize the generalized shift technique from the toolbox to get $z \gg 2^j$, and then use a parallel multiplexer (each partition performs a 2:1 1-bit multiplexer) to get $z \gets \mux_{t_j}(z \gg 2^j, z)$. Algorithm~\ref{alg:parallelFloating:variableShift} shows this in $O(\log(N_x) + {\log}^2(N_x) + N_x)$ steps (with a small constant in $O({\log}^2(N_x) + N_x)$). For bit-parallel \emph{variable normalization}, we also perform $t_j \gets \lnot(z_{N_x-2^j} \lor \cdots \lor z_{N_x-1})$ using the reduction technique from the partition toolbox. This provides bit-parallel variable normalization with the same complexity as bit-parallel variable shift. 

\begin{figure*}[!t]
    \centering
    \includegraphics[width=\linewidth, trim={0cm, 0.3cm, 0cm, 0cm}]{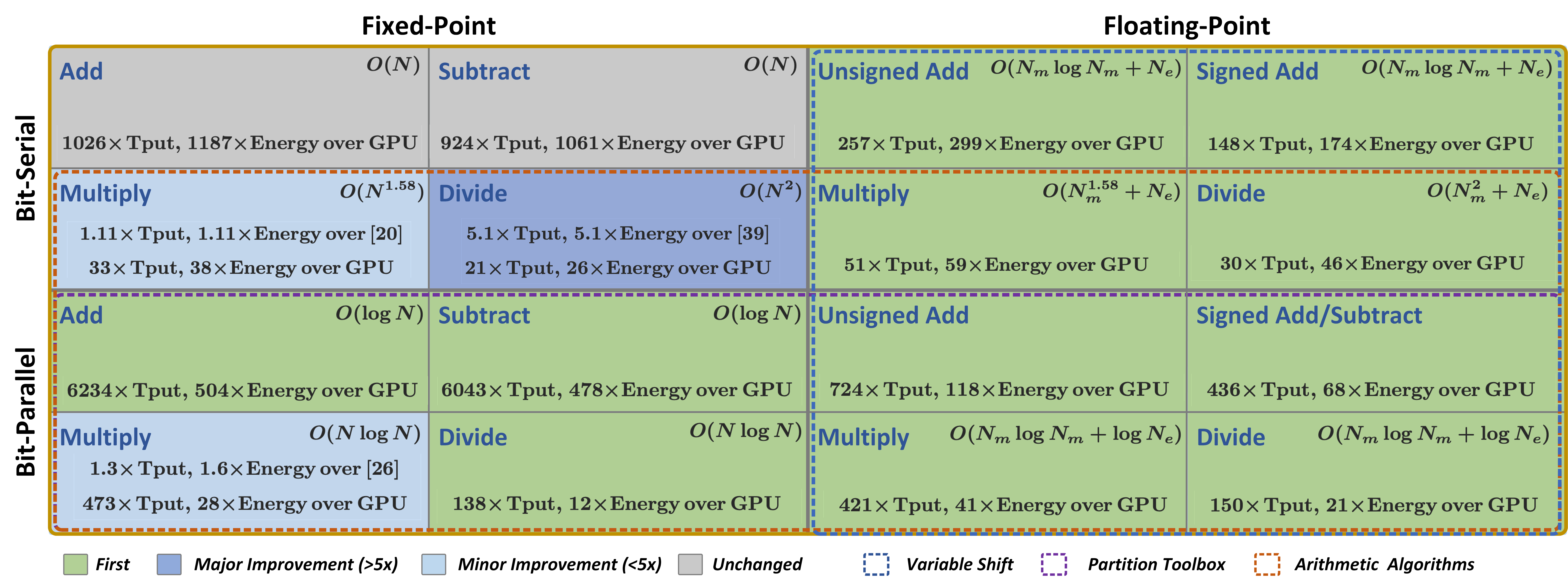}
    \caption{Comparison of AritPIM to both the previous state-of-the-art for PIM (where relevant), and GPUs, for 32-bit numbers. The results compare both Throughput (arithmetic operations per second) and Throughput/Watt (energy). Additional results and details are provided in the repository.}
    \label{fig:results}
    \vspace{-10pt}
\end{figure*}

\section{Evaluation}
\label{sec:results}

We evaluate the AritPIM suite to verify the \emph{correctness} of the algorithms, to \emph{compare} its performance to previous PIM works and alternative solutions (e.g., GPU), and to facilitate its \emph{adoption} by providing an open-access library with comprehensive implementations. The remainder of this section details the evaluation methodology and provides an overview of the results, while focusing primarily on the overall approaches. Further details on the experimental evaluation are available in the README of the code repository\footnote{Available at \url{https://github.com/oleitersdorf/AritPIM}}, alongside implementations for all of the algorithms.

\subsection{Correctness}
\label{sec:results:correctness}

The correctness of the proposed algorithms is verified via a \emph{cycle-accurate} simulation that consists of a PIM simulator and the library of the proposed algorithms. The PIM simulator models a single-row as a binary vector and possesses an interface for performing in-memory gates: a logic gate (e.g., NOR) may be sent to the simulator, and the simulator internally applies the logic gate. The library consists of implementations of the algorithms in this paper, each receiving the parameters of the algorithm (e.g., $N$) and outputting a sequence of in-memory gates. Together, correctness is verified as follows: the inputs are manually written to the PIM simulator's internal memory (e.g., integers $x,y$), a sequence of logic gates is generated by the library and sent to the PIM simulator to be applied internally, and then the output (e.g., $z$) is manually read and compared to the expected value (e.g., $x + y$). For floating-point numbers, we compare to python-based floating-point operations (which adhere to the IEEE-754 standard). For partitions, we adopt the simulator for the minimal-model from PartitionPIM~\cite{PartitionPIM}.

\subsection{Performance Comparison}
\label{sec:results:comparison}

We compare the performance of AritPIM to previous PIM works (where relevant) and to GPU. Figure~\ref{fig:results} summarizes the comparison, demonstrating both significant improvements over previous PIM works (in the few cases where previous works exist) and vast potential for high throughput compared to GPUs. Additional results for different parameters are available in the README of the code repository (e.g., cycle counts, energy, area); this section continues by detailing the methodology that derived these results.

For the comparison to alternative PIM works, we compare the cycle-count of the algorithms when implemented with the same underlying set of logic gates of NOT/NOR\footnote{Previous works were modified to assume $9$ NORs per full adder to provide a fair comparison between the algorithmic concepts. \revision{Specifically, \mbox{\cite{Ameer, GraphLayout}} were upgraded from a 12-NOR full-adder to a 9-NOR full-adder, and the NOR-based implementation of \mbox{\cite{MultPIM}} was adopted.}}. This provides a fair comparison between the algorithms as the underlying conditions are identical, and thus the only differences are the algorithmic concepts.

For the comparison to GPU, we consider a specific potential architecture of PIM and compare to experimental results from a modern GPU. Specifically, we consider a case-study of memristive PIM supporting the NOT/NOR gates~\cite{Nishil, RACER}, with memristor parameters derived from RACER~\cite{RACER}, constructing an 8GB memory from $1024 \times 1024$ crossbars.\footnote{The evaluated vector dimension is $64M$ elements without loss of generality (as there are $64M$ rows in an 8GB memory of $1024\times 1024$ arrays). Yet, AritPIM also supports larger vectors with identical Throughput and Throughput/Watt by using batches (e.g., $128M$-element vector addition is performed by storing two elements per row per vector, and then performing two $64M$-element vector additions serially).} The peripheral correctness and the evaluation of electrical limitations follow \revision{from ongoing works~\mbox{\cite{RACER, Nishil, 9056847}} that explore device and circuit models.} These parameters reflect estimates of memristor performanc~\cite{RACER}, and may differ between technologies; regardless, the overall trends and orders of magnitude remain and the proposed algorithms are directly applicable to all of the different forms of memristive PIM. Therefore, future work may select the most appropriate technology for PIM according to its specific parameters, and use the same algorithms from AritPIM. For GPU, we initialize vectors of $64M$ numbers in the GPU memory and measure the GPU performance  when performing vectored operations (e.g., addition) on those vectors. Specifically, we utilize an NVIDIA RTX 3070 GPU with the \verb|PyTorch|~\cite{PyTorch} profiling tools. Notice that this corresponds to data-intensive scenarios where the data does not fit within the cache; furthermore, we observed that the experimental results are almost identical to the theoretical upper-bound established by the GPU memory throughput -- indicating that the memory wall is indeed the bottleneck. This observation also validates that other computational architectures (e.g., FPGA) subject to the same memory bandwidth will not outperform the observed GPU performance.

\subsection{Adoption}
\label{sec:results:adoption}
One of the goals of this paper is to provide a foundational suite of arithmetic operations to advance PIM towards large-scale applications. Therefore, we have taken the following steps to facilitate the adoption of the proposed algorithms:
\begin{itemize}
    \item \emph{Uniformity:} All of the algorithms conform to the same interface for the inputs and outputs (e.g., $z \gets x \circ y$ for $\circ \in \{+, -, *, \backslash\}$ all conform to the same interface). This simplifies the usage of arithmetic in a PIM architecture towards an application.
    \item \emph{Open-Access Algorithms:} The implementations of the proposed algorithms are publicly available in the code repository. This enables their integration within cycle-accurate simulators for larger applications.
    \item \emph{Verified Results:} We provide verified results for the cycle counts, energy, and area, of all of the algorithms, thereby enabling their usage in a larger application without requiring a cycle-accurate simulator.
\end{itemize}

\section{Conclusion}
\label{sec:conclusion}

As the memory-wall continues to limit the performance of modern computing systems, processing-in-memory (PIM) systems are rethinking the separation of storage and logic units to provide massive parallelism for bitwise logic within the memory itself. This paper extends this bitwise parallelism to high-throughput arithmetic in order to provide a foundation for large-scale PIM applications. We study the four elementary functions for both fixed-point and floating-point numbers, and via two emerging computational approaches of bit-serial and bit-parallel execution -- providing the first algorithms in the literature for a majority of cases. Overall, this paper may be fundamental in the integration of large-scale applications with different PIM technologies.

\vspace{-5pt}

\section*{Acknowledgments}
This work was supported in part by the European Research Council through the European Union's Horizon 2020 Research and Innovation Programme under Grants 757259 and 101069336, and in part by the Israel Science Foundation under Grant 1514/17.

\vspace{-5pt}

\bibliographystyle{IEEEtran}
\bibliography{refs}

\begin{IEEEbiography}[{\includegraphics[width=1in,height=1.25in,clip,keepaspectratio]{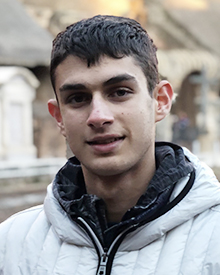}}]{Orian Leitersdorf} (Student Member, IEEE) 
received the B.Sc. degree from the Technion, Haifa in 2022, and is currently a PhD candidate at the Technion, Haifa, Israel. He was a scholar at both the Technion Excellence Program and the Lapidim CS Excellence Program, he previously received the Gutwirth Excellence Scholarship, and he is currently a recipient of the Jacobs Excellence Scholarship. Further, he has received several awards, including the Early Career Researcher Paper Award at the ISITA 2022 conference. His current research aims to advance digital PIM towards fundamental applications (e.g., matrix operations, graph algorithms) while also addressing challenges such as reliability. Further, his research interests also include information theory and constrained coding.
\end{IEEEbiography}

\vspace{-10pt}

\begin{IEEEbiography}[{\includegraphics[width=1in,height=1.25in,clip,keepaspectratio]{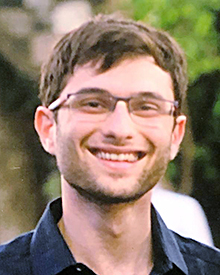}}]{Dean Leitersdorf}
received the B.Sc. and Ph.D. degrees from the Technion, Haifa, in 2019 and 2022. He was a recipient of several awards, including the 2018 Best Student Paper award at OPODIS, 2019 Best Student Paper award at PODC, the 2021 Jacobs Excellent Paper award (Technion), and the 2021 Jacobs Excellence certificate (Technion). His current research interests are in the fields of distributed graph algorithms, focusing on distance computations, sparse matrix multiplication and subgraph existence.
\end{IEEEbiography}

\begin{IEEEbiography}[{\includegraphics[width=1in,height=1.25in,clip,keepaspectratio]{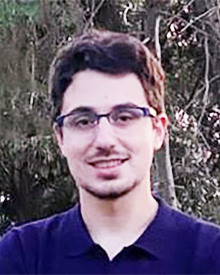}}]{Jonathan Gal} is currently studying towards his B.Sc in Computer Science and Mathematics at the Technion, Haifa, Israel, as part of the Technion Excellence Program. Jonathan Gal won a bronze medal in 2015 in the APIO (Asia Pacific Informatics Olympiad) and the IOI (International Informatics Olympiad). Between 2016 and 2020 he worked as a software engineer, focusing on image processing.
\end{IEEEbiography}

\vspace{22pt}

\begin{IEEEbiography}[{\includegraphics[width=1in,height=1.25in,clip,keepaspectratio]{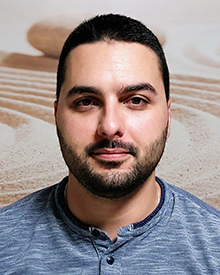}}]{Mor Dahan}
is currently finishing his B.Sc. in Electrical Engineering at the Technion, Haifa, Israel. He joined Intel in 2018 as a DevOps engineer, and since 2020 he has been working as a hardware designer (focusing on pre-silicon verification).
\end{IEEEbiography}

\vspace{25pt}

\begin{IEEEbiography}[{\includegraphics[width=1in,height=1.25in,clip,keepaspectratio]{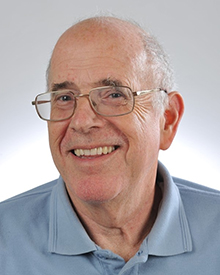}}]{Ronny Ronen} (Fellow, IEEE) 
received the B.Sc. and M.Sc. degrees in computer science from the Technion, Haifa, Israel, in 1978 and 1979, respectively. He is a Senior Researcher with the Andrew and Erna Viterbi Faculty of Electrical \& Computer Engineering at the Technion. He was with Intel Corporation from 1980 to 2017 in various technical and managerial positions. In his last role, he led the Intel Collaborative Research Institute for Computational Intelligence. He was the Director of microarchitecture research and a Senior Staff Computer Architect at the Intel Haifa Development Center until 2011. He led the development of several system software products and tools, including the Intel Pentium processor performance simulator and several compiler efforts. In these roles, he led/was involved in the initial definition and pathfinding of major leading-edge Intel processors. He holds over 80 issued patents and has published over 35 papers.
\end{IEEEbiography}

\vspace{22pt}

\begin{IEEEbiography}[{\includegraphics[width=1in,height=1.25in,clip,keepaspectratio]{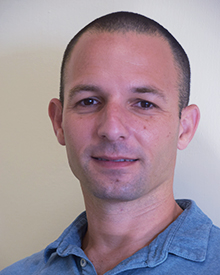}}]{Shahar Kvatinsky}
(Senior Member, IEEE) is an Associate Professor at the Viterbi Faculty of Electrical and Computer Engineering, Technion. Shahar received the B.Sc. degree in Computer Engineering and Applied Physics and an MBA degree in 2009 and 2010, respectively, both from the Hebrew University of Jerusalem, and the Ph.D. degree in Electrical Engineering from the Technion in 2014. From 2006 to 2009, he worked as a circuit designer at Intel. From 2014 and 2015, he was a post-doctoral research fellow at Stanford University. Kvatinsky is a member of the Israel Young Academy. He is the head of the Architecture and Circuits Research Center at the Technion and chair of the IEEE Circuits and Systems in Israel. Kvatinsky has been the recipient of numerous awards including: 2020 MDPI Electronics Young Investigator Award, 2019 Wolf Foundation's Krill Prize, 2015 IEEE Guillemin-Cauer Best Paper Award, ERC starting grant, and the 2017 Pazy Memorial Award.
\end{IEEEbiography}

\end{document}